\documentclass[floatfix,reprint,amsmath,amssymb,prapplied,superscriptaddress,aps]{revtex4-2}
\usepackage[dvipsnames]{xcolor}
\usepackage{graphicx}
\usepackage{dcolumn}
\usepackage{bm}

\usepackage{siunitx}
\usepackage{hyperref}
\hypersetup{
colorlinks=true, breaklinks=true, bookmarks=true,bookmarksnumbered,
urlcolor=RoyalBlue, linkcolor=RoyalBlue, citecolor=RoyalBlue, 
pdftitle={},
pdfauthor={\textcopyright}, 
pdfsubject={},
pdfkeywords={}, 
pdfcreator={pdfLaTeX},
pdfproducer={LaTeX}
}
\begin{document}

\title{The role of magnetic dipolar interactions in skyrmion lattices}
\author{Elizabeth M Jefremovas}
\email{martinel@.uni-mainz.de}
\affiliation{Institute of Physics, Johannes Gutenberg University Mainz, Staudingerweg 7, 55128 Mainz, Germany}
\author{Kilian Leutner}
\affiliation{Institute of Physics, Johannes Gutenberg University Mainz, Staudingerweg 7, 55128 Mainz, Germany}
\author{Miriam G Fischer}
\affiliation{Institute of Physics, Johannes Gutenberg University Mainz, Staudingerweg 7, 55128 Mainz, Germany}
\author{Jorge Marqu\'es--March\'an}
\affiliation{Institute of Materials Sciences of Madrid -- CSIC, 28049 Madrid, Spain}
\author{Thomas B. Winkler}
\affiliation{Institute of Physics, Johannes Gutenberg University Mainz, Staudingerweg 7, 55128 Mainz, Germany}
\author{Agustina Asenjo}
\affiliation{Institute of Materials Sciences of Madrid -- CSIC, 28049 Madrid, Spain}
\author{Robert Fr{\"o}mter}
\affiliation{Institute of Physics, Johannes Gutenberg University Mainz, Staudingerweg 7, 55128 Mainz, Germany}
\author{Jairo Sinova}
\affiliation{Institute of Physics, Johannes Gutenberg University Mainz, Staudingerweg 7, 55128 Mainz, Germany}
\affiliation{Institute of Physics, Czech Academy of Sciences, Cukrovarnick\'a 10, 162 00 Praha 6, Czech Republic}
\author{Mathias Kl{\"a}ui}
\affiliation{Institute of Physics, Johannes Gutenberg University Mainz, Staudingerweg 7, 55128 Mainz, Germany}

\date{\today}

\begin{abstract}
Magnetic skyrmions are promising candidates for information and storage technologies. In the last years, magnetic multilayer systems have been tuned to enable room-temperature skyrmions, stable even in the absence of external magnetic field. There are several models describing the properties of an isolated skyrmion in a homogeneous background for single repetition multilayer stack, however, the description on how the equilibrium skyrmion size in lattices scales with increasing the number of repetitions of the stack remains unaddressed. This question is essential for fundamental and practical perspectives, as the behaviour of an ensemble of skyrmions differs from the isolated case. Based on a multilayer stack hosting a skyrmion lattice, we have carried out a series of imaging experiments scaling up the dipolar interaction by repeating $n$ times the multilayer unit, from $n =1$ up to $n=30$. We have developed an analytical description for the skyrmion radius in the whole multilayer regime, \textit{i.e.}, from thin to thick film limits. Furthermore, we provide insight on how nucleation by an externally applied field can give rise to a lattice with more skyrmions (thus, overfilled) than the predicted by the calculations.
\end{abstract}

\maketitle

\section*{Introduction}

Skyrmions are topologically non-trivial spin textures that act as 2D quasiparticles \cite{bogdanov2020physical, fert2017magnetic}. Their topological stability and low critical current densities for current-induced motion make them promising candidates for energy-efficient information storage and processing technologies \cite{nagaosa2013topological, brems2021circuits, raab2022brownian, song2020skyrmion, leutner2023skyrmion}. First realized in bulk systems, such as chiral magnets \cite{tokura2020magnetic, muhlbauer2009skyrmion}, in the last years, the study of topological spin-textures in magnetic multilayers has gained rising attention, as these allow to tune the magnetic interactions by selecting the material parameters. In this sense, the relevant properties for spin textures, as perpendicular magnetic anisotropy (PMA), Dzyaloshinskii--Moriya  interaction (DMI), interlayer exchange coupling, or dipolar interactions (stray field) can be tuned by selecting the materials forming the multilayer stack. Furthermore, small variations of the thickness of these layers (typically in the sub-nanometer range) can be used to strongly modify the magnetic behaviour of the stack \cite{vedmedenko2019interlayer, kammerbauer2023controlling, yuan2012effect}. This capability to control the magnetic behaviour of the stack by adjusting its composition together with the straightforward integration with today's complementary metal-oxide semiconductor (CMOS) technology, motivates the interest behind magnetic multilayers, constituting an excellent playground to stabilize and study the properties of magnetic skyrmions \cite{dupe2016engineering, soumyanarayanan2017tunable, dohi2022thin, jiang2017skyrmions, satywali2021microwave}. Over the last years, different stack compositions have been reported to host room-temperature stable skyrmions, nucleated at low magnetic fields, and providing ultra-low pinning \cite{zhou2021voltage, moreau2016additive, gruber2023300}. This constitutes already a step forward for the envisaged applications of skyrmions in information technologies, while at the same time, the accessible experimental parameters of such stacks enable fundamental studies of the physics governing the formation of topologically non-trivial spin structure arrangement using in-house techniques, such as Kerr microscopy or Magnetic Force Microscopy (MFM) \cite{raab2022brownian, casiraghi2019individual}. \newline

Whereas DMI has traditionally been employed as the main mechanism for stabilizing spin textures \cite{yuan2016skyrmion, iwasaki2013universal, zhou2014reversible, li2014tailoring, jiang2015blowing}, in the last years, alternative mechanisms have been put forward, as it is the case of, for instance, magnetic frustration \cite{karube2018disordered, von2017enhanced}, or the competition between long-range dipolar interactions, ferromagnetic exchange, and magnetic anisotropy \cite{montoya2017tailoring, desautels2019realization, hrabec2017current, zhang2020formation, heigl2021dipolar}. Besides requiring lower DMI values to stabilize spin textures, in dipolar-stabilized skyrmions, the lateral spin structure size can be signficantly larger than the domain wall width $\Delta$ \cite{ezawa2010giant}, which typically results in significantly larger skyrmions ($\mu$m \textit{vs.} nm) compared to DMI-stabilized ones. This translates into the practical advantage that light-optical experimental techniques can be used for probing the textures, either statically or dynamically under oscillating magnetic fields and/or currents \cite{dohi2019formation, zazvorka2020skyrmion, gruber2022skyrmion, raab2022brownian}, facilitating the experimental studies. In addition to this, magnetic dipolar interactions can easily be tuned by increasing the number of repetitions of the magnetic multilayer unit, providing a wider range of coupling strengths compared to the accessible DMI values, restricted to the material itself, as it is an interfacial effect. Besides the capability of being able to tailor the skyrmion state, a systematic study over the texture-repetition dependence also sets the foundations to understand the emergence of 3-dimensional (3D) spin textures that have recently shown in experiments to critically depend on the number of repetitions of the multilayer unit forming the stack \cite{kent2019generation, kent2021creation, grelier2022three, yu20243d}. \newline

Despite significant interest in the generation and manipulation of skyrmions, there are still fundamental aspects of the skyrmion phase in these multilayers that have not been fully explored. For example, the dependence of the radius $R$ and periodicity $P$ on the material parameters and external fields for stray field-stabilized skyrmion lattices, where the domain wall width $\Delta$ is much smaller than the radius $R$. In this way, there are few recent works describing analytically skyrmion lattices,  mostly addressing the skyrmion arrangement, both for DMI- \cite{wang2018theory} or dipolar-stabilized systems \cite{zazvorka2020skyrmion}, although the majority of the works offer only a qualitative understanding over the skyrmion lattice properties \cite{rohart2013skyrmion, li2019anatomy, vidal2017stability, leonov2016properties}. Furthermore, the dependence of the skyrmion lattice parameters (skyrmion size, skyrmion periodicity) with dipolar interactions remains unheard from an analytical perspective, as the recent reports focus on \textit{isolated} skyrmions, where a single skyrmion can freely expand to minimize the total energy in an infinite system without being affected by the surrounding neighbours \cite{buttner2018theory}. Therefore, these theories are not suitable to describe ensembles of skyrmions, which we will address in the following as \textit{skyrmion lattices}, disregarding the presence or absence of local or long-range order. This constitutes a gap in fundamental knowledge, but also, a major issue towards applications, as for instance, skyrmion reservoirs are conceived to hold the maximum amount of skyrmions \cite{prychynenko2018magnetic, raab2022brownian, sun2023experimental}. The essential difference in this aspect is that we consider the minimum of the energy for a system, where the number (or density) of skyrmions is a free and optimized parameter itself and the skyrmion lattice is lower in energy than the FM state. For the experiment this implies that the nucleation barrier is low and can be overcome using appropriate field protocols. We will show in the following, that dipolar-stabilized skyrmions show the opposite size dependence upon increasing the number of repetitions when they arrange into lattices compared to what has been reported when they are isolated. \newline

To understand the impact of the multilayer repetitions, we present here a systematic experimental study on the variation of skyrmion size as a function of the number of repetitions in an ultra-low pinning skyrmion-lattice system, and develop an analytical model capable of accounting for our findings. While the experiments exhibit qualitatively the same behaviour known for the equilibrium sizes in checkerboard, stripe, or bubble domain patterns \cite{lemesh2017accurate, cape1971magnetic, kaplan1993domain}, we have derived the numerical prefactors that result from the energy minimization, which are different in skyrmion lattices and stripe domain patterns. For the experiments, we have engineered a magnetic multilayer from a Ta/CoFeB-based stack that hosts N\'eel skyrmions, as verified by DMI and current-induced dynamics measurements previously performed in our group ($e.g.$ \cite{gruber2022skyrmion, zazvorka2020skyrmion, raab2022brownian}). Starting from this platform, we have tuned the relative strength of the dipolar interactions by increasing the number of repetitions, $n$, from $n =$1 to 30. A simple sketch of the stack is shown in Fig. \ref{imagenes} \textbf{f)}. We propose an analytical model for the skyrmion size and period as a function of the number of repetitions, which we expand to the limit of $n\rightarrow \infty$ under the assumption of a vertically homogeneous magnetization. Our calculations yield excellent agreement with the experiments, as well as with the results from micromagnetic simulation, enabling us to fully capture the physics underlying these dipolar-stabilized skyrmion lattices. \newline

\section*{Results and discussion}
\subsection*{Experimental observation of skyrmion lattices in multilayer stacks}

\begin{figure}[tb!]
\centering
\resizebox{1.0\columnwidth}{!}{\includegraphics{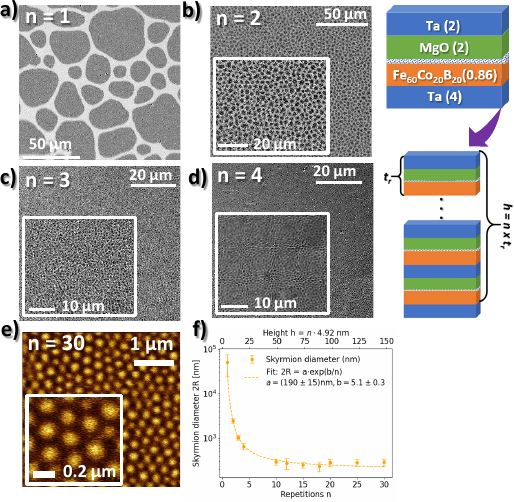}}
\caption{\label{imagenes} a)--d) Kerr-microscopy images corresponding to $n =$ 1 to 4 repetitions of the multilayer stack taken at room temperature (RT) and in zero external field. Light gray represents upwards magnetization ($+z$). A strong decrease of skyrmion size with increasing $n$ is observed. Insets give a higher-magnification view on the skyrmion lattices. e) Magnetic Force Microscopy (MFM) image of the $n =$ 30 repetitions, taken at RT and zero external magnetic field with an ultra-low moment customized tip. Insets show a higher-magnification view on the hexagonal clusters. f) Skyrmion diameter $2R$ \textit{vs.} number of repetitions $n$ as extracted from the images, together with a fit to a phenomenological exponential decay, yielding $2R = a e^{b/n}$, being $a = 190\pm 15$ nm and $b = 5.1\pm 0.3$. The right-hand side sketch provides a schematic view on the stack, with numbers in brackets indicating film thicknesses (in nm). Each unit of thickness $t_{\rm r}$ is repeated $n$ times, yielding the total height $h$ of the stack as $h = n\cdot t_{\rm r}$.}
\end{figure}

The meta-stable skyrmion lattice state is a result of the interplay of several material parameters, namely the saturation magnetization $M_{\rm s}$, the uniaxial magnetic anisotropy $K_{\rm u}$, as well as the DMI strength ($D$) and the exchange stiffness $A$ \cite{lemesh2018current}. To quantitatively determine these parameters, we have used SQUID magnetometry. Within the experimental uncertainty we find constant values of saturation magnetization $M_{\rm s}=$ 600(30) kA/m and $K_{\rm u} = 300(40)$ kJ/m\textsuperscript{3} over the whole series of prepared stacks. From the observation of a constant uniaxial anisotropy for the whole multilayer series, \textit{i.e.}, not significantly affected by the number of repetitions $n$, we imply that $K_{\rm u}$ is also homogeneous over the FM layers of all individual repetition units. With the anisotropy being generally the most susceptible quantity to any structural changes at the interfaces, it indicates that also the interfacial DMI keeps homogeneous over the whole stack, \textit{i.e.}, will not change with $n$. The same holds for the exchange stiffness in an unaltered local atomic environment of the CoFeB layers. We assume therefore values of $D = 0.9$ mJ/m$^{2}$ and $A = 10$ pJ/m, in good agreement with values reported in literature for similar stack compositions \cite{zazvorka2019thermal, casiraghi2019individual, ge2023constructing}. The contributions to the total energy that originate from $A$, $D$, and $K_{\rm u}$ scale linearly with $n$, so the skyrmion size resulting from a minimization of these energy terms alone would not change with $n$. The dipolar energy depending on $M_{\rm s}$, in contrast, also has contributions from the mutual interactions between all layers and will thus scale differently with the number of repetitions $n$. So by changing $n$ we can selectively tune the effect of dipolar interactions, giving rise to changes in skyrmion size. \newline

Fig.$~$\ref{imagenes}a)--d) show Kerr microscopy images of $n = $ 1 to 4 repetitions measured at zero field, while \textbf{e)} and \textbf{f)} include $n =$15 and 30, measured with MFM. The skyrmion size of $n =$10--30, below the space resolution of optical-based techniques, calls for using such scanning probe techiques. All images show disordered arrays of almost-circular skyrmions, but the most striking observation is related to the conspicuous decrease of the skyrmion size as the number of repetition increases. As shown in Fig. \ref{imagenes} \textbf{a)}, the $n =$ 1 repetition hosts $\approx$ 40 $ \mu$m skyrmions, with a significant size variation of about $50 \%$, while the skyrmions at $n=$ 2 are already a factor of 20 smaller ($\approx$ 2.3 $ \mu$m). At $n=4$ the average skyrmion diameter is 520 nm, already close to the resolution limit of the Kerr microscope. This is why the higher repetitions (from $n = $ 10 to 30) have been probed with MFM, capable of sensing forces at the atomic-level. Fig. \ref{imagenes} \textbf{e)} includes an representative image of the $n = 30$ stack, where even some hexagonal clusters of skyrmions (see inset) are formed. The decrease of the skyrmion size with increasing dipolar coupling fits to a phenomenologically-motivated exponential decay [see Fig. \ref{imagenes}\textbf{f)}], agrees with the analytical model for checkerboard and stripe domains proposed by Kaplan and Gehring \cite{kaplan1993domain}.\newline

In order to get a first understanding of our experimental results, where a systematic decrease of the skyrmion size upon increasing dipolar coupling is measured, on the contrary to the case of isolated skyrmions \cite{buttner2018theory}, we hypothesize that the existence of neighbouring skyrmions result in an effective skyrmion-skyrmion interaction mediated by dipolar interactions and this is an interaction term that is not accounted for in the isolated skyrmion model. This skyrmion-skyrmion interaction would result effectively in an additional stray field for the single skyrmion under consideration, acting as an additional magnetic field. As a result of such field, the skyrmion size decreases compared to the isolated situation. Our hypothesis becomes particularly evident in the case of $n =$1 [see Fig. \ref{imagenes} \textbf{a)}], where the skyrmions show a broad size distribution and a particular arrangement, with the large skyrmions being surrounded by groups of smaller skyrmions. This indicates a very shallow energy minimum of the total energy as a function of the skyrmion size as a result of the inhomogeneous stray field. Upon increasing $n$, the stray field becomes more homogeneous, yielding a better defined energy minimum and thus, a more homogeneous skyrmion size. To test our hypothesis, we have used the model proposed by B{\"u}ttner \textit{et al.}, including \textit{ad hoc} an external magnetic field. Our experimental results are reproduced by a Zeeman field in the range of $\mu$T, in line with our hypothesis of a field generated by the neighbouring skyrmions, motivating our development of a full analytical model which includes this field contribution from the other skyrmions to account for the skyrmion-skyrmion interaction, resulting in a model where both the skyrmion size and the lattice periodicity are taken into account to minimize the energy of the system. \newline

\subsection*{Analytical model for dipolar-stabilized skyrmion lattices}

In order to capture the physics of skyrmion lattices, we have developed an analytical model, capable of providing an analytical description for the skyrmion size and periodicity. Our model, that treats the full stack as a single homogeneous magnetic layer (\textit{effective medium model}  \cite{woo2016observation}), assumes skyrmions as rigid circular cylinders which extend across the layers, with a magnetization configuration centered at the origin \cite{buttner2017field, romming2015field, boulle2016room}. Moreover, for a lattice of skyrmions, the energy $E$ of one unit cell can be decomposed into \cite{cape1971magnetic}

\begin{equation} \label{energias}
    E = E_{\rm DW} + E_{\rm d, film} + E_{\rm d, c} + E_{\rm d, c\leftrightarrow film} + E_{\rm d, c \leftrightarrow c} 
\end{equation}

being $E_{\rm DW}$ the domain wall energy of the skyrmion (including volume charges), $E_{\rm d, film}$ the self--energy (demagnetization energy) of the ferromagnetic film, $E_{\rm d, c}$ the self--energy of an isolated cylinder, $E_{\rm d, c\leftrightarrow film}$ the interaction energy between the cylinder and the ferromagnetic film due to the demagnetization field, and $E_{\rm d, c \leftrightarrow c}$ the interaction between one cylinder and the cylinders in adjacent unit cells. The main innovation of our proposed model concerns the first and the last term in the previous equation, $i.e.$, $E_{\rm DW}$ and $E_{\rm d, c \leftrightarrow c}$. Thus, we will restrict our discussions here to these two terms, being the remaining ones calculated following the analytical description of isolated skyrmions by B{\"u}ttner \textit{et al.} \cite{buttner2018theory}. \newline

Starting with $E_{\rm DW}$, this term includes exchange, anisotropy and DMI interactions; the volume charges consequence of the demagnetizing field; and the contribution from the surface charges due to the inhomogeniety of the domain wall itself. In addition to this, an offset to account for the non--zero value of the anisotropy in the uniform areas, $\Sigma$, is included. Therefore, this term reads

\begin{equation}\label{domain_wall_energy}
    \tilde{E}_{\rm DW} = 2\pi R \frac{\sigma(h)}{\mu_{0}M_{\rm s}^{2}}h + P^{2} h \Sigma
\end{equation}
where $\tilde{E}$ is the normalized energy $E$ by a factor $\frac{E}{\mu_{0}M_{\rm s}^{2}}$, and the domain wall energy density

\begin{equation}\label{domain_wall_density}
    \frac{\sigma(h)}{\mu_{0}M_{\rm s}^{2}} = \frac{\sigma_{AKD}}{\mu_{0}M_{\rm s}^{2}} + \frac{\sigma_{d}(h)}{h}
\end{equation}

being $\sigma_{\rm AKD}$ the domain wall width arising from exchange, anisotropy and DMI, which is obtained from micromagnetic simulations , and $\sigma_{\rm d}(h)$ the functional form of the domain wall energy density for the demagnetizing field. Note that this expression does not depend explicitly anymore on the domain wall width, obtained as $\Delta \approx 11\,\mathrm{nm}$, as it is implicit through the minimization process. The obtained value agrees with the one reported for similar systems \cite{buttner2017field, buttner2018theory}, and matches the $R/\Delta>>1$ criterion for N\'eel-type skyrmions. \newline

To calculate the $E_{\rm cylinder \leftrightarrow cylinder}$ term, we have considered two steps. First, the interaction energy between two cylinders is calculated via a multipole expansion of the surface contribution from the demagnetizing energy. Then, a two-dimensional sum of this single cylinder-cylinder interaction is solved using the Euler-Maclaurin formula \cite{kac2002euler}. This way, the interaction from one skyrmion with all the surrounding skyrmions in the 2D lattice, is fully captured. \newline

This way, the radius $R_0$ and periodicity $P_0$ of the energy minima is given by $\partial_P (E(R,P)/P^2)=0$ and $\partial_R (E(R,P)/P^2)=0$, with $E$ given by Eq. \ref{energias}. For the case where no external field is applied, a factor 50\% skyrmion and 50\% background can be used for the interaction potential of the skyrmions considered here in good approximation (also verified with simulations). This delivers the relation $P=R \sqrt{2\pi}$ for the square lattice and {$P=R \sqrt[4]{\frac{16\pi^{2}}{3}} $} for the hexagonal lattice. Accordingly, explicit analytical solutions for the two limits of $h \rightarrow 0$ (thin film), and $h \rightarrow \infty$ (thick film regime) can be written as

\begin{equation}\label{formulon}
    R(h) =\left\{ \begin{array}{lr} 1.025 \text{ }h \text{  exp}[\frac{\pi\lambda}{h}]  & \text{if } h \rightarrow 0\text{ and } \frac{P}{h} \rightarrow \infty \\ \\ 
    2.53 \text{ }\sqrt{\lambda h} & \text{if } h \rightarrow \infty\text{ and } \frac{P}{h} \rightarrow 0 \end{array} \right.
\end{equation}

where $\lambda = \frac{\sigma_{\rm AKD}}{\mu_{0}M_{\rm s}}^{2}$. The lower limit in Eq. \ref{formulon} reproduces the same scaling as the one already described for stripe domains \cite{lemesh2017accurate, kaplan1993domain, cape1971magnetic}, with a different prefactor for the case of skyrmion lattices. The upper limit, which accounts for an increase of the skyrmion size at large number of repetitions, reproduces the same scaling of the Kittel model for one--dimensional domain walls \cite{kittel1946theory}. \newline

\begin{figure}[tb!]
\centering
\resizebox{1.0\columnwidth}{!}{\includegraphics{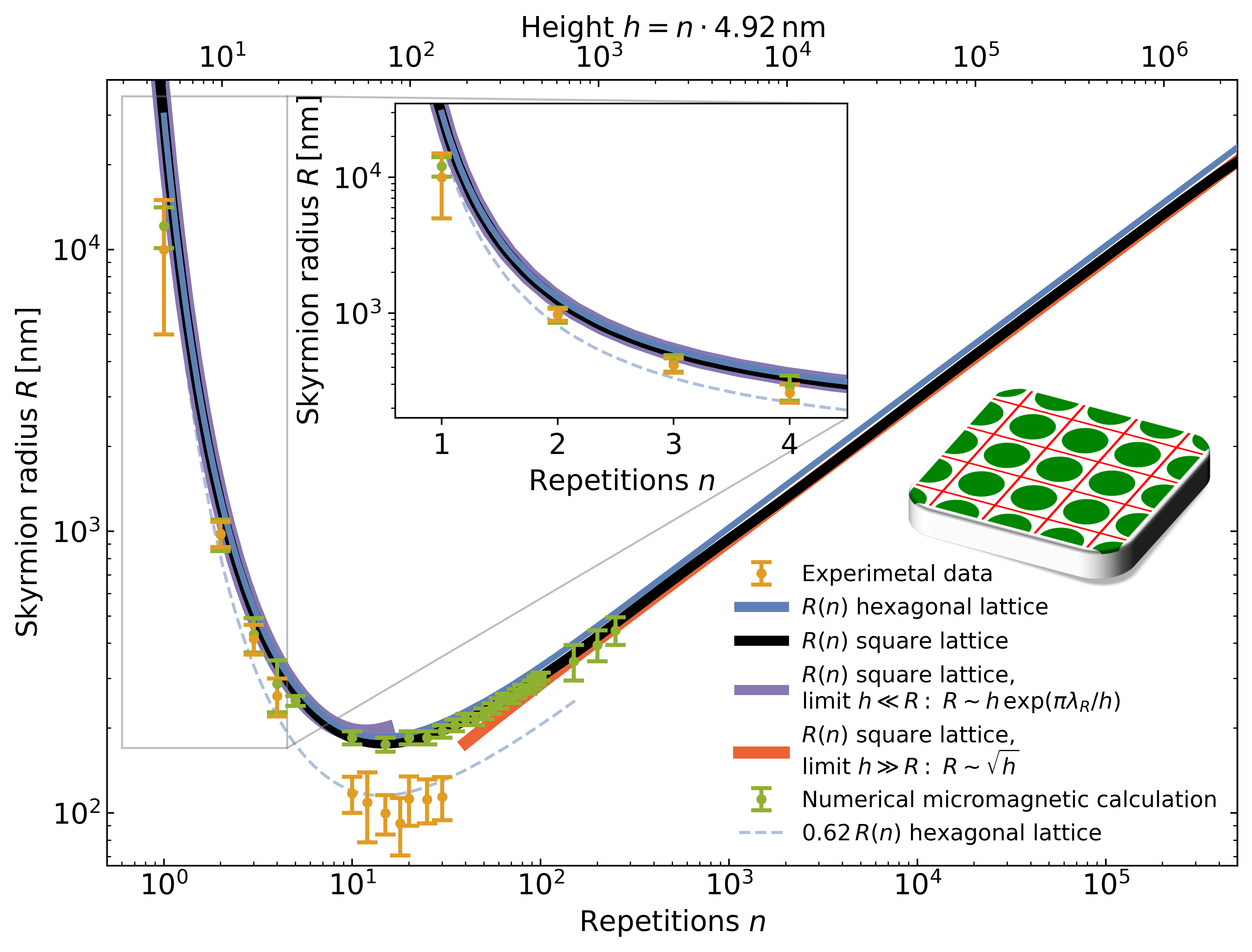}}
\caption{\label{analytical} Skyrmion radius $R$ \textit{vs.} number of repetitions $n$: experimental and numerically simulated data points, together with the analytical model. No significant discrepancy is found between the square and hexagonal arrangements. The sketch represents the lattice and the skyrmions (in green), for a square--lattice arrangement. Inset zooms-in the $n=$1-4 range, where the agreement with the experimental values is excellent. A discrepancy with the experimental values between $n=$10-30 (factor 2 smaller than the predictions) is observed. However, our analytical model can reproduce the tendency of the experimental data values (see dashed line), with a fit of a prefactor (0.62).}
\end{figure}

Fig. \ref{analytical} includes the prediction from our model, together with the experimental results shown in Fig. \ref{imagenes}. We have also simulated numerically values for the skyrmion sizes (green points in Fig. \ref{analytical}). Importantly, we have simulated the system with the effective medium approach (\textit{effective ferromagnetic model} \cite{woo2016observation}), and compared our results to the simulation of the full layers, without the simplification described by Woo \textit{et al.} \cite{woo2016observation}. In both cases, a reasonable agreement with out model is obtained (discrepancy around 10\%), guaranteeing the validity of the effective medium approach, which reduces the computational costs. As observed in Fig. \ref{analytical}, the model reproduces very well the decrease of the skyrmion size as a function of the number of repetitions in the low $n$ regime, where the agreement between experiments, simulations and analytical model is perfect. For the upper limit $n\rightarrow \infty$, an increase of the skyrmion size is predicted, similar to the Kittel model for domain walls \cite{kittel1946theory}, and fitting our simulation results. \newline

However, there is a discrepancy in the \textit{plateau} regime, from $n= 10$ to 30, where a factor $\approx$2 difference between the prediction and the experimental size is obtained. The reason for this discrepancy will be discussed in the next section. In particular, we need to take into account the formation of the spin structure that can result in metastable states while theoretically we consider the lowest energy state. \newline

\subsection*{Mechanism underlying stripes to skyrmion transition}

In Fig. \ref{analytical}, a discrepancy between the numerical and experimental data is found in the \textit{plateau regime} separating the low and high repetitions \textit{i.e.}, from $n=$ 10-30, where the experimental values are systematically a factor $\approx 2$ smaller than the calculated ones. We have evaluated several facts that could lead to such a deviation. First, we considered the possibility of deviations from the skyrmion tube shape across the multilayer, leading to a decrease in the skyrmion size at the top and bottom surfaces, as discussed in a recent work by T. Srivastava \textit{et al.} for a $n=$ 20 repetitions FeCoB-based stack \cite{srivastava2023resonant}. However, none of our simulations (with or without the effective medium approach) showed significant deformations. Our simulations also ruled out the hybrid skyrmion scenario, with a N\'eel-Bloch-N\'eel type across the stack \cite{liyanage2023three}. Second, pinning effects can also be excluded, as i) the presence of hexagonal clusters in $n =$30 [(see inset in \textbf{e)}] serves as an indication that a deep local energy minima of the system is reached; and ii) our experimental results show a reduced skyrmion size than expected, whereas pinning effects shall lead to larger skyrmions close to those pinning sites. Finally, we checked the influence of the tip magnetic field on the sample, and determined it is not a critical factor. A remaining possible explanation for the factor 2 is to analyze carefully the nucleation procedure that drives the magnetization from stripes to skyrmions. \newline

\begin{figure*}[tb!]
\centering
\resizebox{2\columnwidth}{!}{\includegraphics{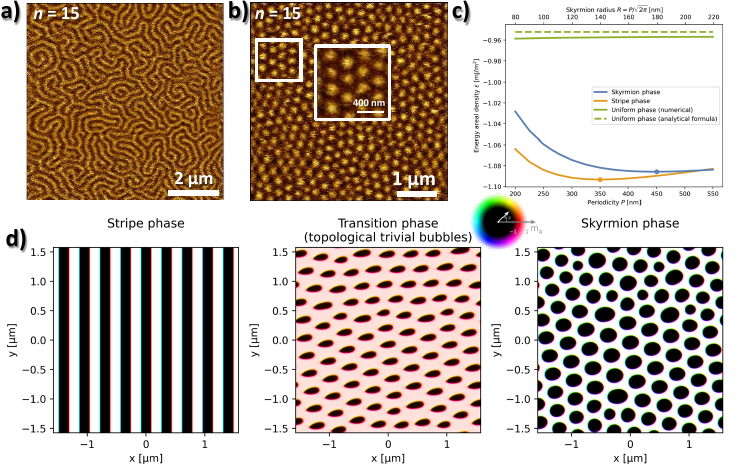}}
\caption{\label{transition} Stripe to skyrmion transition: a) and b) include representative MFM images of $n=$ 15 after OOP saturation and IP saturation, respectively. Note the difference between the multidomain pattern and the skyrmion state. In c), the areal energy density is plotted for the periodicity ranges between 250 to 500 nm, which correspond to the experimental values of the higher repetitions ($n=$ 10--30). The stripe phase is the lowest energy state for the experimental periodicities. The bottom panel d) shows the simulations of the full process from stripe-to-skyrmion phases mediated via a transition phase which consists of topological trivial bubbles. The configurations were obtained by minimizing the total energy of the system, the field was linearly increased/decreased up to/from the maximum field $\mu_{0}H_{max}$ = (150 mT, 0, 20 mT). The magnetic contrast is described by the color wheel.}
\end{figure*}

Fig. \ref{transition} \textbf{a)} includes MFM images corresponding to the $n =$ 15 sample after applying 135 mT OOP, which is enough to saturate the stacks, and the magnetization relaxes back to a stripe state when the field is removed. Note the maze-like pattern, that holds for all repetitions, which is an indication of the low-pinning of the samples \cite{hubert1998magnetic}. The periodicity extracted from the MFM images (2D-FFT) corresponds to (300 $\pm$ 30) nm for all $n =$ 10-30. When the same field (135 mT) is applied IP with a small OOP component (below 20 mT), the magnetization undergoes a transition to a skyrmion lattice phase, as shown in Fig. \ref{transition} \textbf{b)}. Note the formation of hexagonal clusters of skyrmions (see inset), which is a good indication of low-pinning. Fig. \ref{transition} \textbf{c)} includes the calculated energy areal density corresponding to both stripe and skyrmion phases. Both energy minima are close in energy, and well below the one corresponding to the uniform phase. Note however that, while the calculated energy minima for the stripe phase (350 nm in $n =$15) is close to our experimental result (290 $\pm$ 20) nm, there is a factor 2 of difference between the calculated energy minima for skyrmions (450 nm) and the experimental value (250 $\pm$ 40) nm. This factor 2 difference, already observed in the analytical model presented in the previous section (Fig.~\ref{analytical}), is explored next taking into account the nucleation process of the skyrmion lattice. \newline

Based on micromagnetic simulations, we propose a mechanism that enables the transition from the stripe to the skyrmion phase through an intermediate bubble phase. Starting from a multidomain situation with the domains 100\% parallel [\textit{stripes}, see Fig. \ref{transition} \textbf{d)}], applying a IP field (150 mT) at a 90$^{\circ} \pm 15^{\circ}$ relative to the stripe alignment with a small OOP component (20 mT, which corresponds to $\approx~6^{\circ}$ OOP) deforms the domains, which break up forming an intermediate phase, consisting on skyrmions with zero topological charge, referred to here as \textit{bubbles}. Such \textit{bubbles} exhibit N\'eel and Bloch walls, in contrast to skyrmion--antiskyrmion droplets recently reported \cite{sisodia2021chiral}. According to our simulations, the bubble phase is not stable, and it evolves to the final metastable skyrmion energy phase with topological charge $Q = 1$ when the field is removed. The resulting nucleated skyrmion phase is an overfilled skyrmion lattice in a local meta-stable minimum, thus, not the global energy minima calculated and shown in Fig. \ref{transition} \textbf{c)}. This has a crucial impact to the lattice periodicity, as the overfilled skyrmion lattice has a much lower periodicity (calculated values between 250-300 nm for $n =~$10-30) than the 50-50 skyrmion phase (50\% skyrmions-50\% uniform state), for which a value of $P \approx$ 450 nm is calculated. The $P$ of the overfilled skyrmion lattice matches the experimental values, accounting for the factor 2 difference between the analytical model (that assumes 50\% skyrmions-50\% uniform state) and the experimental results for $n =~$10-30 (overfilled skyrmion lattice, 70\% skyrmions-30\% uniform state). On the contrary, in the case of the low repetitions ($n =~$ 1-4), the experimentally obtained skyrmion lattice corresponds to the the actual global energy minima, (50-50, no overfilling). Bearing in mind the fact that the periodicity of the skyrmion state highly depends on the starting stripe periodicity, being the energy minima of both stripe and skyrmion phases very close for these 1-4 repetitions, the transition is accomplished without going through an overfilled lattice. It is also worth it mentioning that we are capable of observing such a striking factor 2 difference in the n = 10-30 repetitions because of the steeper energy minima (larger curvature of the energy landscape) compared to $n =~$1-4, which implies that deviations from the calculated periodicity are more noticeable. \newline

\section*{Conclusions}

In this work, we have systematically studied the effect of dipolar interactions in the stabilization of skyrmion lattices, elucidating the role of both the skyrmion size and lattice periodicity for the minimization of the energy system. Our results demonstrate the opposite scaling with dipolar coupling with respect to isolated skyrmions, enabling a way to increase skyrmion density by increasing the dipolar coupling, which also favours the robustness of the spin textures against external magnetic fields. Note that, at the lower limit ($n \rightarrow$0), our model reproduces the same scaling as the one already described for stripe domains \cite{lemesh2017accurate, kaplan1993domain, cape1971magnetic}, with a different prefactor, while the upper limit ($n\rightarrow \infty$) reproduces the same scaling of the Kittel model for one--dimensional domain walls \cite{kittel1946theory}. We also describe the nucleation procedure of the skyrmion lattice starting from a stripe situation, elucidating the key role of the starting stripe periodicity (\textit{i.e.}, energy minima) in the final skyrmion lattice state (overfilled lattice $vs.$ no overfilled). \newline

Besides their intrinsic fundamental interest, our results are relevant for future skyrmion--based devices. The precise understanding on how the skyrmion--skyrmion interaction alters the minimum energy state of the system is essential to control skyrmion reservoirs \cite{pinna2020reservoir, prychynenko2018magnetic, leutner2023skyrmion} and skymion dynamics \cite{iwasaki2013current, litzius2020role, kim2017current}. Particularly critical is the case of the reported transition from stripes to skyrmions, achieved at room temperature and magnetic fields in the mT regime, which could potentially be used as a simple mechanism to nucleate skyrmion vya the in--plane rotation of the magnetic field, but also, to increased storage density by reaching an overfilled skyrmion lattice. In addition to this, the extremely sensitive size decrease of the skyrmion size at low dipolar coupling (\textit{i.e.}, $n =$1-4) could be used as an indicator to monitor the dipolar coupling in Magnetic-Tunnel-Junctions (MTJ) or current-induced motion of skyrmions in devices (\textit{e.g.} race-track memories), where inhomogenities in the dipolar field, caused by the interactions with the different components, would be identified by monitoring the skyrmion size.\newline

\section*{Methods}

Magnetic multilayer stacks of the form Ta(4)/[Co\textsubscript{20}Fe\textsubscript{60}B\textsubscript{20}(0.86)/Ta(0.06)/MgO(2)/Ta(2)]$\times n$ (layer thicknesses in nm, 0.01 nm precision), with $n$ being the number of repetitions, were deposited using a Singulus Rotaris magnetron sputtering tool with a base pressure of $3 \times 10^{-8}$ mbar onto Si/SiO\textsubscript{2} substrates. The surface roughness of the films was checked by Atomic Force Microscopy (AFM) to be below 1 nm. To reduce the strong PMA induced at the interface between Co\textsubscript{20}Fe\textsubscript{60}B\textsubscript{20} and MgO, and thus enable skyrmion nucleation at RT, a dusting Ta layer is introduced in between. The randomly distributed Ta atoms weaken the Fe--O and Co--O bonds, reducing the PMA of the stack \cite{yu2016room}. The increase in dipolar energy density has been achieved by increasing the number of repetitions $n$.  \newline

Hysteresis $M$\textit{vs.}$H$ loops were performed using a superconducting quantum interference device (SQUID) for both in-plane and out-of-plane configurations at room temperature and for fields between -2 and 2 T. \newline

Magnetic contrast was established using the polar magneto-optical Kerr effect (MOKE) in a commercially available Kerr microscope from Evico Magnetics GmbH. Electromagnetic coils allow to apply simultaneously out-of-plane (OOP) and in-plane (IP) fields up to 13 mT and 120 mT, respectively. The microscope is set up in a thermally stabilized flow box, ensuring constant 292K temperature conditions. Data were acquired by a CCD camera as gray-scale images, with a field of view of 200 × 150 $\mu$m$^{2}$. Skyrmions were nucleated in a constant OOP field, from $\approx 100\; \mu$T for $n =$ 1 to $\approx 6$ mT for $n=$ 4, by applying a saturating IP field pulse of $\approx 0.5$ s duration, in the same way as in \cite{zazvorka2020skyrmion, gruber2023300}. All images shown and discussed in the present work are taken at zero field (both IP and OOP). Skyrmion sizes were determined from the images by machine-learning-based detection using U-Net \cite{labrie2024machine} and cross-checked by ImageJ \cite{collins2007imagej}. \newline

MFM and the corresponding AFM images were recorded using a Nanotec scanning probe microscopy system controlled by WSxM software \cite{horcas2007wsxm}. In order to minimize the influence of the tip magnetic moment on the sample, customized ultra-low moment tips were produced. For this, a thin film of Co ($\approx$ 10 nm) was deposited onto the front face of a commercial Nanosensors PPP-FMR tip. This ultra-low coating thickness \cite{lopez2022high} ensured a proficient signal-to-noise ratio during the measurements, carried out in amplitude modulation mode. \newline

Micromagnetic simulations were performed using Mumax3 \cite{vansteenkiste2014design}, and verified with MicMag2 \cite{winkler2021skyrmion, de2016multiscale, woo2016observation, prychynenko2018magnetic, MicMag2}. The simulations were performed also by treating the multilayer stack as a single-layer homogeneous ferromagnet (\textit{effective ferromagnet}) \cite{woo2016observation}, scaling the $M_{s}$, $A$, $K_{eff}$ and $D$ to effective parameters by the ratio $t_{m}$/$t_{r}$, being $t_{m}$ the material thickness per repetition and $t_{r}$ the thickness of one repetition. A discretization of $\Delta x = \Delta y = 5$ nm was employed. Both square and hexagonal lattice arrangements, with periodicity $P = l_{x} = l_{y}$ for the former, and $P = l_{x} = \frac{l_{y}}{\sqrt{3}}$ for the latter, were tested, including periodic boundary conditions in both $l_{x}$ and $l_{y}$ directions. \newline

\bibliography{main}

\begin{thebibliography}{78}%
\makeatletter
\providecommand \@ifxundefined [1]{%
 \@ifx{#1\undefined}
}%
\providecommand \@ifnum [1]{%
 \ifnum #1\expandafter \@firstoftwo
 \else \expandafter \@secondoftwo
 \fi
}%
\providecommand \@ifx [1]{%
 \ifx #1\expandafter \@firstoftwo
 \else \expandafter \@secondoftwo
 \fi
}%
\providecommand \natexlab [1]{#1}%
\providecommand \enquote  [1]{``#1''}%
\providecommand \bibnamefont  [1]{#1}%
\providecommand \bibfnamefont [1]{#1}%
\providecommand \citenamefont [1]{#1}%
\providecommand \href@noop [0]{\@secondoftwo}%
\providecommand \href [0]{\begingroup \@sanitize@url \@href}%
\providecommand \@href[1]{\@@startlink{#1}\@@href}%
\providecommand \@@href[1]{\endgroup#1\@@endlink}%
\providecommand \@sanitize@url [0]{\catcode `\\12\catcode `\$12\catcode `\&12\catcode `\#12\catcode `\^12\catcode `\_12\catcode `\%12\relax}%
\providecommand \@@startlink[1]{}%
\providecommand \@@endlink[0]{}%
\providecommand \url  [0]{\begingroup\@sanitize@url \@url }%
\providecommand \@url [1]{\endgroup\@href {#1}{\urlprefix }}%
\providecommand \urlprefix  [0]{URL }%
\providecommand \Eprint [0]{\href }%
\providecommand \doibase [0]{https://doi.org/}%
\providecommand \selectlanguage [0]{\@gobble}%
\providecommand \bibinfo  [0]{\@secondoftwo}%
\providecommand \bibfield  [0]{\@secondoftwo}%
\providecommand \translation [1]{[#1]}%
\providecommand \BibitemOpen [0]{}%
\providecommand \bibitemStop [0]{}%
\providecommand \bibitemNoStop [0]{.\EOS\space}%
\providecommand \EOS [0]{\spacefactor3000\relax}%
\providecommand \BibitemShut  [1]{\csname bibitem#1\endcsname}%
\let\auto@bib@innerbib\@empty
\bibitem [{\citenamefont {Bogdanov}\ and\ \citenamefont {Panagopoulos}(2020)}]{bogdanov2020physical}%
  \BibitemOpen
  \bibfield  {author} {\bibinfo {author} {\bibfnamefont {A.~N.}\ \bibnamefont {Bogdanov}}\ and\ \bibinfo {author} {\bibfnamefont {C.}~\bibnamefont {Panagopoulos}},\ }\href@noop {} {\bibfield  {journal} {\bibinfo  {journal} {Nature Reviews Physics}\ }\textbf {\bibinfo {volume} {2}},\ \bibinfo {pages} {492} (\bibinfo {year} {2020})}\BibitemShut {NoStop}%
\bibitem [{\citenamefont {Fert}\ \emph {et~al.}(2017)\citenamefont {Fert}, \citenamefont {Reyren},\ and\ \citenamefont {Cros}}]{fert2017magnetic}%
  \BibitemOpen
  \bibfield  {author} {\bibinfo {author} {\bibfnamefont {A.}~\bibnamefont {Fert}}, \bibinfo {author} {\bibfnamefont {N.}~\bibnamefont {Reyren}},\ and\ \bibinfo {author} {\bibfnamefont {V.}~\bibnamefont {Cros}},\ }\href@noop {} {\bibfield  {journal} {\bibinfo  {journal} {Nature Reviews Materials}\ }\textbf {\bibinfo {volume} {2}},\ \bibinfo {pages} {1} (\bibinfo {year} {2017})}\BibitemShut {NoStop}%
\bibitem [{\citenamefont {Nagaosa}\ and\ \citenamefont {Tokura}(2013)}]{nagaosa2013topological}%
  \BibitemOpen
  \bibfield  {author} {\bibinfo {author} {\bibfnamefont {N.}~\bibnamefont {Nagaosa}}\ and\ \bibinfo {author} {\bibfnamefont {Y.}~\bibnamefont {Tokura}},\ }\href@noop {} {\bibfield  {journal} {\bibinfo  {journal} {Nature nanotechnology}\ }\textbf {\bibinfo {volume} {8}},\ \bibinfo {pages} {899} (\bibinfo {year} {2013})}\BibitemShut {NoStop}%
\bibitem [{\citenamefont {Brems}\ \emph {et~al.}(2021)\citenamefont {Brems}, \citenamefont {Kl{\"a}ui},\ and\ \citenamefont {Virnau}}]{brems2021circuits}%
  \BibitemOpen
  \bibfield  {author} {\bibinfo {author} {\bibfnamefont {M.~A.}\ \bibnamefont {Brems}}, \bibinfo {author} {\bibfnamefont {M.}~\bibnamefont {Kl{\"a}ui}},\ and\ \bibinfo {author} {\bibfnamefont {P.}~\bibnamefont {Virnau}},\ }\href@noop {} {\bibfield  {journal} {\bibinfo  {journal} {Applied Physics Letters}\ }\textbf {\bibinfo {volume} {119}} (\bibinfo {year} {2021})}\BibitemShut {NoStop}%
\bibitem [{\citenamefont {Raab}\ \emph {et~al.}(2022)\citenamefont {Raab}, \citenamefont {Brems}, \citenamefont {Beneke}, \citenamefont {Dohi}, \citenamefont {Roth{\"o}rl}, \citenamefont {Kammerbauer}, \citenamefont {Mentink},\ and\ \citenamefont {Kl{\"a}ui}}]{raab2022brownian}%
  \BibitemOpen
  \bibfield  {author} {\bibinfo {author} {\bibfnamefont {K.}~\bibnamefont {Raab}}, \bibinfo {author} {\bibfnamefont {M.~A.}\ \bibnamefont {Brems}}, \bibinfo {author} {\bibfnamefont {G.}~\bibnamefont {Beneke}}, \bibinfo {author} {\bibfnamefont {T.}~\bibnamefont {Dohi}}, \bibinfo {author} {\bibfnamefont {J.}~\bibnamefont {Roth{\"o}rl}}, \bibinfo {author} {\bibfnamefont {F.}~\bibnamefont {Kammerbauer}}, \bibinfo {author} {\bibfnamefont {J.~H.}\ \bibnamefont {Mentink}},\ and\ \bibinfo {author} {\bibfnamefont {M.}~\bibnamefont {Kl{\"a}ui}},\ }\href@noop {} {\bibfield  {journal} {\bibinfo  {journal} {Nature Communications}\ }\textbf {\bibinfo {volume} {13}},\ \bibinfo {pages} {6982} (\bibinfo {year} {2022})}\BibitemShut {NoStop}%
\bibitem [{\citenamefont {Song}\ \emph {et~al.}(2020)\citenamefont {Song}, \citenamefont {Jeong}, \citenamefont {Pan}, \citenamefont {Zhang}, \citenamefont {Xia}, \citenamefont {Cha}, \citenamefont {Park}, \citenamefont {Kim}, \citenamefont {Finizio}, \citenamefont {Raabe} \emph {et~al.}}]{song2020skyrmion}%
  \BibitemOpen
  \bibfield  {author} {\bibinfo {author} {\bibfnamefont {K.~M.}\ \bibnamefont {Song}}, \bibinfo {author} {\bibfnamefont {J.-S.}\ \bibnamefont {Jeong}}, \bibinfo {author} {\bibfnamefont {B.}~\bibnamefont {Pan}}, \bibinfo {author} {\bibfnamefont {X.}~\bibnamefont {Zhang}}, \bibinfo {author} {\bibfnamefont {J.}~\bibnamefont {Xia}}, \bibinfo {author} {\bibfnamefont {S.}~\bibnamefont {Cha}}, \bibinfo {author} {\bibfnamefont {T.-E.}\ \bibnamefont {Park}}, \bibinfo {author} {\bibfnamefont {K.}~\bibnamefont {Kim}}, \bibinfo {author} {\bibfnamefont {S.}~\bibnamefont {Finizio}}, \bibinfo {author} {\bibfnamefont {J.}~\bibnamefont {Raabe}}, \emph {et~al.},\ }\href@noop {} {\bibfield  {journal} {\bibinfo  {journal} {Nature Electronics}\ }\textbf {\bibinfo {volume} {3}},\ \bibinfo {pages} {148} (\bibinfo {year} {2020})}\BibitemShut {NoStop}%
\bibitem [{\citenamefont {Leutner}\ \emph {et~al.}(2023)\citenamefont {Leutner}, \citenamefont {Winkler}, \citenamefont {Gruber}, \citenamefont {Fr{\"o}mter}, \citenamefont {G{\"u}ttinger}, \citenamefont {Fangohr},\ and\ \citenamefont {Kl{\"a}ui}}]{leutner2023skyrmion}%
  \BibitemOpen
  \bibfield  {author} {\bibinfo {author} {\bibfnamefont {K.}~\bibnamefont {Leutner}}, \bibinfo {author} {\bibfnamefont {T.~B.}\ \bibnamefont {Winkler}}, \bibinfo {author} {\bibfnamefont {R.}~\bibnamefont {Gruber}}, \bibinfo {author} {\bibfnamefont {R.}~\bibnamefont {Fr{\"o}mter}}, \bibinfo {author} {\bibfnamefont {J.}~\bibnamefont {G{\"u}ttinger}}, \bibinfo {author} {\bibfnamefont {H.}~\bibnamefont {Fangohr}},\ and\ \bibinfo {author} {\bibfnamefont {M.}~\bibnamefont {Kl{\"a}ui}},\ }\href@noop {} {\bibfield  {journal} {\bibinfo  {journal} {Physical Review Applied}\ }\textbf {\bibinfo {volume} {20}},\ \bibinfo {pages} {064021} (\bibinfo {year} {2023})}\BibitemShut {NoStop}%
\bibitem [{\citenamefont {Tokura}\ and\ \citenamefont {Kanazawa}(2020)}]{tokura2020magnetic}%
  \BibitemOpen
  \bibfield  {author} {\bibinfo {author} {\bibfnamefont {Y.}~\bibnamefont {Tokura}}\ and\ \bibinfo {author} {\bibfnamefont {N.}~\bibnamefont {Kanazawa}},\ }\href@noop {} {\bibfield  {journal} {\bibinfo  {journal} {Chemical Reviews}\ }\textbf {\bibinfo {volume} {121}},\ \bibinfo {pages} {2857} (\bibinfo {year} {2020})}\BibitemShut {NoStop}%
\bibitem [{\citenamefont {M{\"u}hlbauer}\ \emph {et~al.}(2009)\citenamefont {M{\"u}hlbauer}, \citenamefont {Binz}, \citenamefont {Jonietz}, \citenamefont {Pfleiderer}, \citenamefont {Rosch}, \citenamefont {Neubauer}, \citenamefont {Georgii},\ and\ \citenamefont {B{\"o}ni}}]{muhlbauer2009skyrmion}%
  \BibitemOpen
  \bibfield  {author} {\bibinfo {author} {\bibfnamefont {S.}~\bibnamefont {M{\"u}hlbauer}}, \bibinfo {author} {\bibfnamefont {B.}~\bibnamefont {Binz}}, \bibinfo {author} {\bibfnamefont {F.}~\bibnamefont {Jonietz}}, \bibinfo {author} {\bibfnamefont {C.}~\bibnamefont {Pfleiderer}}, \bibinfo {author} {\bibfnamefont {A.}~\bibnamefont {Rosch}}, \bibinfo {author} {\bibfnamefont {A.}~\bibnamefont {Neubauer}}, \bibinfo {author} {\bibfnamefont {R.}~\bibnamefont {Georgii}},\ and\ \bibinfo {author} {\bibfnamefont {P.}~\bibnamefont {B{\"o}ni}},\ }\href@noop {} {\bibfield  {journal} {\bibinfo  {journal} {Science}\ }\textbf {\bibinfo {volume} {323}},\ \bibinfo {pages} {915} (\bibinfo {year} {2009})}\BibitemShut {NoStop}%
\bibitem [{\citenamefont {Vedmedenko}\ \emph {et~al.}(2019)\citenamefont {Vedmedenko}, \citenamefont {Riego}, \citenamefont {Arregi},\ and\ \citenamefont {Berger}}]{vedmedenko2019interlayer}%
  \BibitemOpen
  \bibfield  {author} {\bibinfo {author} {\bibfnamefont {E.~Y.}\ \bibnamefont {Vedmedenko}}, \bibinfo {author} {\bibfnamefont {P.}~\bibnamefont {Riego}}, \bibinfo {author} {\bibfnamefont {J.~A.}\ \bibnamefont {Arregi}},\ and\ \bibinfo {author} {\bibfnamefont {A.}~\bibnamefont {Berger}},\ }\href@noop {} {\bibfield  {journal} {\bibinfo  {journal} {Physical review letters}\ }\textbf {\bibinfo {volume} {122}},\ \bibinfo {pages} {257202} (\bibinfo {year} {2019})}\BibitemShut {NoStop}%
\bibitem [{\citenamefont {Kammerbauer}\ \emph {et~al.}(2023)\citenamefont {Kammerbauer}, \citenamefont {Choi}, \citenamefont {Freimuth}, \citenamefont {Lee}, \citenamefont {Fr{\"o}mter}, \citenamefont {Han}, \citenamefont {Lavrijsen}, \citenamefont {Swagten}, \citenamefont {Mokrousov},\ and\ \citenamefont {Kl{\"a}ui}}]{kammerbauer2023controlling}%
  \BibitemOpen
  \bibfield  {author} {\bibinfo {author} {\bibfnamefont {F.}~\bibnamefont {Kammerbauer}}, \bibinfo {author} {\bibfnamefont {W.-Y.}\ \bibnamefont {Choi}}, \bibinfo {author} {\bibfnamefont {F.}~\bibnamefont {Freimuth}}, \bibinfo {author} {\bibfnamefont {K.}~\bibnamefont {Lee}}, \bibinfo {author} {\bibfnamefont {R.}~\bibnamefont {Fr{\"o}mter}}, \bibinfo {author} {\bibfnamefont {D.-S.}\ \bibnamefont {Han}}, \bibinfo {author} {\bibfnamefont {R.}~\bibnamefont {Lavrijsen}}, \bibinfo {author} {\bibfnamefont {H.~J.}\ \bibnamefont {Swagten}}, \bibinfo {author} {\bibfnamefont {Y.}~\bibnamefont {Mokrousov}},\ and\ \bibinfo {author} {\bibfnamefont {M.}~\bibnamefont {Kl{\"a}ui}},\ }\href@noop {} {\bibfield  {journal} {\bibinfo  {journal} {Nano Letters}\ }\textbf {\bibinfo {volume} {23}},\ \bibinfo {pages} {7070} (\bibinfo {year} {2023})}\BibitemShut {NoStop}%
\bibitem [{\citenamefont {Yuan}\ \emph {et~al.}(2012)\citenamefont {Yuan}, \citenamefont {Lin}, \citenamefont {Mei}, \citenamefont {Hsu},\ and\ \citenamefont {Kuo}}]{yuan2012effect}%
  \BibitemOpen
  \bibfield  {author} {\bibinfo {author} {\bibfnamefont {F.-T.}\ \bibnamefont {Yuan}}, \bibinfo {author} {\bibfnamefont {Y.-H.}\ \bibnamefont {Lin}}, \bibinfo {author} {\bibfnamefont {J.}~\bibnamefont {Mei}}, \bibinfo {author} {\bibfnamefont {J.-H.}\ \bibnamefont {Hsu}},\ and\ \bibinfo {author} {\bibfnamefont {P.}~\bibnamefont {Kuo}},\ }\href@noop {} {\bibfield  {journal} {\bibinfo  {journal} {Journal of Applied Physics}\ }\textbf {\bibinfo {volume} {111}} (\bibinfo {year} {2012})}\BibitemShut {NoStop}%
\bibitem [{\citenamefont {Dup{\'e}}\ \emph {et~al.}(2016)\citenamefont {Dup{\'e}}, \citenamefont {Bihlmayer}, \citenamefont {B{\"o}ttcher}, \citenamefont {Bl{\"u}gel},\ and\ \citenamefont {Heinze}}]{dupe2016engineering}%
  \BibitemOpen
  \bibfield  {author} {\bibinfo {author} {\bibfnamefont {B.}~\bibnamefont {Dup{\'e}}}, \bibinfo {author} {\bibfnamefont {G.}~\bibnamefont {Bihlmayer}}, \bibinfo {author} {\bibfnamefont {M.}~\bibnamefont {B{\"o}ttcher}}, \bibinfo {author} {\bibfnamefont {S.}~\bibnamefont {Bl{\"u}gel}},\ and\ \bibinfo {author} {\bibfnamefont {S.}~\bibnamefont {Heinze}},\ }\href@noop {} {\bibfield  {journal} {\bibinfo  {journal} {Nature communications}\ }\textbf {\bibinfo {volume} {7}},\ \bibinfo {pages} {11779} (\bibinfo {year} {2016})}\BibitemShut {NoStop}%
\bibitem [{\citenamefont {Soumyanarayanan}\ \emph {et~al.}(2017)\citenamefont {Soumyanarayanan}, \citenamefont {Raju}, \citenamefont {Gonzalez~Oyarce}, \citenamefont {Tan}, \citenamefont {Im}, \citenamefont {Petrovi{\'c}}, \citenamefont {Ho}, \citenamefont {Khoo}, \citenamefont {Tran}, \citenamefont {Gan} \emph {et~al.}}]{soumyanarayanan2017tunable}%
  \BibitemOpen
  \bibfield  {author} {\bibinfo {author} {\bibfnamefont {A.}~\bibnamefont {Soumyanarayanan}}, \bibinfo {author} {\bibfnamefont {M.}~\bibnamefont {Raju}}, \bibinfo {author} {\bibfnamefont {A.}~\bibnamefont {Gonzalez~Oyarce}}, \bibinfo {author} {\bibfnamefont {A.~K.}\ \bibnamefont {Tan}}, \bibinfo {author} {\bibfnamefont {M.-Y.}\ \bibnamefont {Im}}, \bibinfo {author} {\bibfnamefont {A.~P.}\ \bibnamefont {Petrovi{\'c}}}, \bibinfo {author} {\bibfnamefont {P.}~\bibnamefont {Ho}}, \bibinfo {author} {\bibfnamefont {K.}~\bibnamefont {Khoo}}, \bibinfo {author} {\bibfnamefont {M.}~\bibnamefont {Tran}}, \bibinfo {author} {\bibfnamefont {C.}~\bibnamefont {Gan}}, \emph {et~al.},\ }\href@noop {} {\bibfield  {journal} {\bibinfo  {journal} {Nature materials}\ }\textbf {\bibinfo {volume} {16}},\ \bibinfo {pages} {898} (\bibinfo {year} {2017})}\BibitemShut {NoStop}%
\bibitem [{\citenamefont {Dohi}\ \emph {et~al.}(2022)\citenamefont {Dohi}, \citenamefont {Reeve},\ and\ \citenamefont {Kl{\"a}ui}}]{dohi2022thin}%
  \BibitemOpen
  \bibfield  {author} {\bibinfo {author} {\bibfnamefont {T.}~\bibnamefont {Dohi}}, \bibinfo {author} {\bibfnamefont {R.~M.}\ \bibnamefont {Reeve}},\ and\ \bibinfo {author} {\bibfnamefont {M.}~\bibnamefont {Kl{\"a}ui}},\ }\href@noop {} {\bibfield  {journal} {\bibinfo  {journal} {Annual Review of Condensed Matter Physics}\ }\textbf {\bibinfo {volume} {13}},\ \bibinfo {pages} {73} (\bibinfo {year} {2022})}\BibitemShut {NoStop}%
\bibitem [{\citenamefont {Jiang}\ \emph {et~al.}(2017)\citenamefont {Jiang}, \citenamefont {Chen}, \citenamefont {Liu}, \citenamefont {Zang}, \citenamefont {Te~Velthuis},\ and\ \citenamefont {Hoffmann}}]{jiang2017skyrmions}%
  \BibitemOpen
  \bibfield  {author} {\bibinfo {author} {\bibfnamefont {W.}~\bibnamefont {Jiang}}, \bibinfo {author} {\bibfnamefont {G.}~\bibnamefont {Chen}}, \bibinfo {author} {\bibfnamefont {K.}~\bibnamefont {Liu}}, \bibinfo {author} {\bibfnamefont {J.}~\bibnamefont {Zang}}, \bibinfo {author} {\bibfnamefont {S.~G.}\ \bibnamefont {Te~Velthuis}},\ and\ \bibinfo {author} {\bibfnamefont {A.}~\bibnamefont {Hoffmann}},\ }\href@noop {} {\bibfield  {journal} {\bibinfo  {journal} {Physics Reports}\ }\textbf {\bibinfo {volume} {704}},\ \bibinfo {pages} {1} (\bibinfo {year} {2017})}\BibitemShut {NoStop}%
\bibitem [{\citenamefont {Satywali}\ \emph {et~al.}(2021)\citenamefont {Satywali}, \citenamefont {Kravchuk}, \citenamefont {Pan}, \citenamefont {Raju}, \citenamefont {He}, \citenamefont {Ma}, \citenamefont {Petrovi{\'c}}, \citenamefont {Garst},\ and\ \citenamefont {Panagopoulos}}]{satywali2021microwave}%
  \BibitemOpen
  \bibfield  {author} {\bibinfo {author} {\bibfnamefont {B.}~\bibnamefont {Satywali}}, \bibinfo {author} {\bibfnamefont {V.~P.}\ \bibnamefont {Kravchuk}}, \bibinfo {author} {\bibfnamefont {L.}~\bibnamefont {Pan}}, \bibinfo {author} {\bibfnamefont {M.}~\bibnamefont {Raju}}, \bibinfo {author} {\bibfnamefont {S.}~\bibnamefont {He}}, \bibinfo {author} {\bibfnamefont {F.}~\bibnamefont {Ma}}, \bibinfo {author} {\bibfnamefont {A.~P.}\ \bibnamefont {Petrovi{\'c}}}, \bibinfo {author} {\bibfnamefont {M.}~\bibnamefont {Garst}},\ and\ \bibinfo {author} {\bibfnamefont {C.}~\bibnamefont {Panagopoulos}},\ }\href@noop {} {\bibfield  {journal} {\bibinfo  {journal} {Nature communications}\ }\textbf {\bibinfo {volume} {12}},\ \bibinfo {pages} {1909} (\bibinfo {year} {2021})}\BibitemShut {NoStop}%
\bibitem [{\citenamefont {Zhou}\ \emph {et~al.}(2021)\citenamefont {Zhou}, \citenamefont {Mansell},\ and\ \citenamefont {van Dijken}}]{zhou2021voltage}%
  \BibitemOpen
  \bibfield  {author} {\bibinfo {author} {\bibfnamefont {Y.}~\bibnamefont {Zhou}}, \bibinfo {author} {\bibfnamefont {R.}~\bibnamefont {Mansell}},\ and\ \bibinfo {author} {\bibfnamefont {S.}~\bibnamefont {van Dijken}},\ }\href@noop {} {\bibfield  {journal} {\bibinfo  {journal} {Applied Physics Letters}\ }\textbf {\bibinfo {volume} {118}} (\bibinfo {year} {2021})}\BibitemShut {NoStop}%
\bibitem [{\citenamefont {Moreau-Luchaire}\ \emph {et~al.}(2016)\citenamefont {Moreau-Luchaire}, \citenamefont {Moutafis}, \citenamefont {Reyren}, \citenamefont {Sampaio}, \citenamefont {Vaz}, \citenamefont {Van~Horne}, \citenamefont {Bouzehouane}, \citenamefont {Garcia}, \citenamefont {Deranlot}, \citenamefont {Warnicke} \emph {et~al.}}]{moreau2016additive}%
  \BibitemOpen
  \bibfield  {author} {\bibinfo {author} {\bibfnamefont {C.}~\bibnamefont {Moreau-Luchaire}}, \bibinfo {author} {\bibfnamefont {C.}~\bibnamefont {Moutafis}}, \bibinfo {author} {\bibfnamefont {N.}~\bibnamefont {Reyren}}, \bibinfo {author} {\bibfnamefont {J.}~\bibnamefont {Sampaio}}, \bibinfo {author} {\bibfnamefont {C.}~\bibnamefont {Vaz}}, \bibinfo {author} {\bibfnamefont {N.}~\bibnamefont {Van~Horne}}, \bibinfo {author} {\bibfnamefont {K.}~\bibnamefont {Bouzehouane}}, \bibinfo {author} {\bibfnamefont {K.}~\bibnamefont {Garcia}}, \bibinfo {author} {\bibfnamefont {C.}~\bibnamefont {Deranlot}}, \bibinfo {author} {\bibfnamefont {P.}~\bibnamefont {Warnicke}}, \emph {et~al.},\ }\href@noop {} {\bibfield  {journal} {\bibinfo  {journal} {Nature nanotechnology}\ }\textbf {\bibinfo {volume} {11}},\ \bibinfo {pages} {444} (\bibinfo {year} {2016})}\BibitemShut {NoStop}%
\bibitem [{\citenamefont {Gruber}\ \emph {et~al.}(2023)\citenamefont {Gruber}, \citenamefont {Brems}, \citenamefont {Roth{\"o}rl}, \citenamefont {Sparmann}, \citenamefont {Schmitt}, \citenamefont {Kononenko}, \citenamefont {Kammerbauer}, \citenamefont {Syskaki}, \citenamefont {Farago}, \citenamefont {Virnau} \emph {et~al.}}]{gruber2023300}%
  \BibitemOpen
  \bibfield  {author} {\bibinfo {author} {\bibfnamefont {R.}~\bibnamefont {Gruber}}, \bibinfo {author} {\bibfnamefont {M.~A.}\ \bibnamefont {Brems}}, \bibinfo {author} {\bibfnamefont {J.}~\bibnamefont {Roth{\"o}rl}}, \bibinfo {author} {\bibfnamefont {T.}~\bibnamefont {Sparmann}}, \bibinfo {author} {\bibfnamefont {M.}~\bibnamefont {Schmitt}}, \bibinfo {author} {\bibfnamefont {I.}~\bibnamefont {Kononenko}}, \bibinfo {author} {\bibfnamefont {F.}~\bibnamefont {Kammerbauer}}, \bibinfo {author} {\bibfnamefont {M.-A.}\ \bibnamefont {Syskaki}}, \bibinfo {author} {\bibfnamefont {O.}~\bibnamefont {Farago}}, \bibinfo {author} {\bibfnamefont {P.}~\bibnamefont {Virnau}}, \emph {et~al.},\ }\href@noop {} {\bibfield  {journal} {\bibinfo  {journal} {Advanced Materials}\ }\textbf {\bibinfo {volume} {35}},\ \bibinfo {pages} {2208922} (\bibinfo {year} {2023})}\BibitemShut {NoStop}%
\bibitem [{\citenamefont {Casiraghi}\ \emph {et~al.}(2019)\citenamefont {Casiraghi}, \citenamefont {Corte-Le{\'o}n}, \citenamefont {Vafaee}, \citenamefont {Garcia-Sanchez}, \citenamefont {Durin}, \citenamefont {Pasquale}, \citenamefont {Jakob}, \citenamefont {Kl{\"a}ui},\ and\ \citenamefont {Kazakova}}]{casiraghi2019individual}%
  \BibitemOpen
  \bibfield  {author} {\bibinfo {author} {\bibfnamefont {A.}~\bibnamefont {Casiraghi}}, \bibinfo {author} {\bibfnamefont {H.}~\bibnamefont {Corte-Le{\'o}n}}, \bibinfo {author} {\bibfnamefont {M.}~\bibnamefont {Vafaee}}, \bibinfo {author} {\bibfnamefont {F.}~\bibnamefont {Garcia-Sanchez}}, \bibinfo {author} {\bibfnamefont {G.}~\bibnamefont {Durin}}, \bibinfo {author} {\bibfnamefont {M.}~\bibnamefont {Pasquale}}, \bibinfo {author} {\bibfnamefont {G.}~\bibnamefont {Jakob}}, \bibinfo {author} {\bibfnamefont {M.}~\bibnamefont {Kl{\"a}ui}},\ and\ \bibinfo {author} {\bibfnamefont {O.}~\bibnamefont {Kazakova}},\ }\href@noop {} {\bibfield  {journal} {\bibinfo  {journal} {Communications Physics}\ }\textbf {\bibinfo {volume} {2}},\ \bibinfo {pages} {145} (\bibinfo {year} {2019})}\BibitemShut {NoStop}%
\bibitem [{\citenamefont {Yuan}\ and\ \citenamefont {Wang}(2016)}]{yuan2016skyrmion}%
  \BibitemOpen
  \bibfield  {author} {\bibinfo {author} {\bibfnamefont {H.}~\bibnamefont {Yuan}}\ and\ \bibinfo {author} {\bibfnamefont {X.}~\bibnamefont {Wang}},\ }\href@noop {} {\bibfield  {journal} {\bibinfo  {journal} {Scientific reports}\ }\textbf {\bibinfo {volume} {6}},\ \bibinfo {pages} {22638} (\bibinfo {year} {2016})}\BibitemShut {NoStop}%
\bibitem [{\citenamefont {Iwasaki}\ \emph {et~al.}(2013{\natexlab{a}})\citenamefont {Iwasaki}, \citenamefont {Mochizuki},\ and\ \citenamefont {Nagaosa}}]{iwasaki2013universal}%
  \BibitemOpen
  \bibfield  {author} {\bibinfo {author} {\bibfnamefont {J.}~\bibnamefont {Iwasaki}}, \bibinfo {author} {\bibfnamefont {M.}~\bibnamefont {Mochizuki}},\ and\ \bibinfo {author} {\bibfnamefont {N.}~\bibnamefont {Nagaosa}},\ }\href@noop {} {\bibfield  {journal} {\bibinfo  {journal} {Nature communications}\ }\textbf {\bibinfo {volume} {4}},\ \bibinfo {pages} {1463} (\bibinfo {year} {2013}{\natexlab{a}})}\BibitemShut {NoStop}%
\bibitem [{\citenamefont {Zhou}\ and\ \citenamefont {Ezawa}(2014)}]{zhou2014reversible}%
  \BibitemOpen
  \bibfield  {author} {\bibinfo {author} {\bibfnamefont {Y.}~\bibnamefont {Zhou}}\ and\ \bibinfo {author} {\bibfnamefont {M.}~\bibnamefont {Ezawa}},\ }\href@noop {} {\bibfield  {journal} {\bibinfo  {journal} {Nature communications}\ }\textbf {\bibinfo {volume} {5}},\ \bibinfo {pages} {4652} (\bibinfo {year} {2014})}\BibitemShut {NoStop}%
\bibitem [{\citenamefont {Li}\ \emph {et~al.}(2014)\citenamefont {Li}, \citenamefont {Tan}, \citenamefont {Moon}, \citenamefont {Doran}, \citenamefont {Marcus}, \citenamefont {Young}, \citenamefont {Arenholz}, \citenamefont {Ma}, \citenamefont {Yang}, \citenamefont {Hwang} \emph {et~al.}}]{li2014tailoring}%
  \BibitemOpen
  \bibfield  {author} {\bibinfo {author} {\bibfnamefont {J.}~\bibnamefont {Li}}, \bibinfo {author} {\bibfnamefont {A.}~\bibnamefont {Tan}}, \bibinfo {author} {\bibfnamefont {K.}~\bibnamefont {Moon}}, \bibinfo {author} {\bibfnamefont {A.}~\bibnamefont {Doran}}, \bibinfo {author} {\bibfnamefont {M.}~\bibnamefont {Marcus}}, \bibinfo {author} {\bibfnamefont {A.}~\bibnamefont {Young}}, \bibinfo {author} {\bibfnamefont {E.}~\bibnamefont {Arenholz}}, \bibinfo {author} {\bibfnamefont {S.}~\bibnamefont {Ma}}, \bibinfo {author} {\bibfnamefont {R.}~\bibnamefont {Yang}}, \bibinfo {author} {\bibfnamefont {C.}~\bibnamefont {Hwang}}, \emph {et~al.},\ }\href@noop {} {\bibfield  {journal} {\bibinfo  {journal} {Nature communications}\ }\textbf {\bibinfo {volume} {5}},\ \bibinfo {pages} {4704} (\bibinfo {year} {2014})}\BibitemShut {NoStop}%
\bibitem [{\citenamefont {Jiang}\ \emph {et~al.}(2015)\citenamefont {Jiang}, \citenamefont {Upadhyaya}, \citenamefont {Zhang}, \citenamefont {Yu}, \citenamefont {Jungfleisch}, \citenamefont {Fradin}, \citenamefont {Pearson}, \citenamefont {Tserkovnyak}, \citenamefont {Wang}, \citenamefont {Heinonen} \emph {et~al.}}]{jiang2015blowing}%
  \BibitemOpen
  \bibfield  {author} {\bibinfo {author} {\bibfnamefont {W.}~\bibnamefont {Jiang}}, \bibinfo {author} {\bibfnamefont {P.}~\bibnamefont {Upadhyaya}}, \bibinfo {author} {\bibfnamefont {W.}~\bibnamefont {Zhang}}, \bibinfo {author} {\bibfnamefont {G.}~\bibnamefont {Yu}}, \bibinfo {author} {\bibfnamefont {M.~B.}\ \bibnamefont {Jungfleisch}}, \bibinfo {author} {\bibfnamefont {F.~Y.}\ \bibnamefont {Fradin}}, \bibinfo {author} {\bibfnamefont {J.~E.}\ \bibnamefont {Pearson}}, \bibinfo {author} {\bibfnamefont {Y.}~\bibnamefont {Tserkovnyak}}, \bibinfo {author} {\bibfnamefont {K.~L.}\ \bibnamefont {Wang}}, \bibinfo {author} {\bibfnamefont {O.}~\bibnamefont {Heinonen}}, \emph {et~al.},\ }\href@noop {} {\bibfield  {journal} {\bibinfo  {journal} {Science}\ }\textbf {\bibinfo {volume} {349}},\ \bibinfo {pages} {283} (\bibinfo {year} {2015})}\BibitemShut {NoStop}%
\bibitem [{\citenamefont {Karube}\ \emph {et~al.}(2018)\citenamefont {Karube}, \citenamefont {White}, \citenamefont {Morikawa}, \citenamefont {Dewhurst}, \citenamefont {Cubitt}, \citenamefont {Kikkawa}, \citenamefont {Yu}, \citenamefont {Tokunaga}, \citenamefont {Arima}, \citenamefont {R{\o}nnow} \emph {et~al.}}]{karube2018disordered}%
  \BibitemOpen
  \bibfield  {author} {\bibinfo {author} {\bibfnamefont {K.}~\bibnamefont {Karube}}, \bibinfo {author} {\bibfnamefont {J.~S.}\ \bibnamefont {White}}, \bibinfo {author} {\bibfnamefont {D.}~\bibnamefont {Morikawa}}, \bibinfo {author} {\bibfnamefont {C.~D.}\ \bibnamefont {Dewhurst}}, \bibinfo {author} {\bibfnamefont {R.}~\bibnamefont {Cubitt}}, \bibinfo {author} {\bibfnamefont {A.}~\bibnamefont {Kikkawa}}, \bibinfo {author} {\bibfnamefont {X.}~\bibnamefont {Yu}}, \bibinfo {author} {\bibfnamefont {Y.}~\bibnamefont {Tokunaga}}, \bibinfo {author} {\bibfnamefont {T.-h.}\ \bibnamefont {Arima}}, \bibinfo {author} {\bibfnamefont {H.~M.}\ \bibnamefont {R{\o}nnow}}, \emph {et~al.},\ }\href@noop {} {\bibfield  {journal} {\bibinfo  {journal} {Science advances}\ }\textbf {\bibinfo {volume} {4}},\ \bibinfo {pages} {eaar7043} (\bibinfo {year} {2018})}\BibitemShut {NoStop}%
\bibitem [{\citenamefont {von Malottki}\ \emph {et~al.}(2017)\citenamefont {von Malottki}, \citenamefont {Dup{\'e}}, \citenamefont {Bessarab}, \citenamefont {Delin},\ and\ \citenamefont {Heinze}}]{von2017enhanced}%
  \BibitemOpen
  \bibfield  {author} {\bibinfo {author} {\bibfnamefont {S.}~\bibnamefont {von Malottki}}, \bibinfo {author} {\bibfnamefont {B.}~\bibnamefont {Dup{\'e}}}, \bibinfo {author} {\bibfnamefont {P.~F.}\ \bibnamefont {Bessarab}}, \bibinfo {author} {\bibfnamefont {A.}~\bibnamefont {Delin}},\ and\ \bibinfo {author} {\bibfnamefont {S.}~\bibnamefont {Heinze}},\ }\href@noop {} {\bibfield  {journal} {\bibinfo  {journal} {Scientific reports}\ }\textbf {\bibinfo {volume} {7}},\ \bibinfo {pages} {12299} (\bibinfo {year} {2017})}\BibitemShut {NoStop}%
\bibitem [{\citenamefont {Montoya}\ \emph {et~al.}(2017)\citenamefont {Montoya}, \citenamefont {Couture}, \citenamefont {Chess}, \citenamefont {Lee}, \citenamefont {Kent}, \citenamefont {Henze}, \citenamefont {Sinha}, \citenamefont {Im}, \citenamefont {Kevan}, \citenamefont {Fischer} \emph {et~al.}}]{montoya2017tailoring}%
  \BibitemOpen
  \bibfield  {author} {\bibinfo {author} {\bibfnamefont {S.}~\bibnamefont {Montoya}}, \bibinfo {author} {\bibfnamefont {S.}~\bibnamefont {Couture}}, \bibinfo {author} {\bibfnamefont {J.}~\bibnamefont {Chess}}, \bibinfo {author} {\bibfnamefont {J.}~\bibnamefont {Lee}}, \bibinfo {author} {\bibfnamefont {N.}~\bibnamefont {Kent}}, \bibinfo {author} {\bibfnamefont {D.}~\bibnamefont {Henze}}, \bibinfo {author} {\bibfnamefont {S.}~\bibnamefont {Sinha}}, \bibinfo {author} {\bibfnamefont {M.-Y.}\ \bibnamefont {Im}}, \bibinfo {author} {\bibfnamefont {S.}~\bibnamefont {Kevan}}, \bibinfo {author} {\bibfnamefont {P.}~\bibnamefont {Fischer}}, \emph {et~al.},\ }\href@noop {} {\bibfield  {journal} {\bibinfo  {journal} {Physical Review B}\ }\textbf {\bibinfo {volume} {95}},\ \bibinfo {pages} {024415} (\bibinfo {year} {2017})}\BibitemShut {NoStop}%
\bibitem [{\citenamefont {Desautels}\ \emph {et~al.}(2019)\citenamefont {Desautels}, \citenamefont {DeBeer-Schmitt}, \citenamefont {Montoya}, \citenamefont {Borchers}, \citenamefont {Je}, \citenamefont {Tang}, \citenamefont {Im}, \citenamefont {Fitzsimmons}, \citenamefont {Fullerton},\ and\ \citenamefont {Gilbert}}]{desautels2019realization}%
  \BibitemOpen
  \bibfield  {author} {\bibinfo {author} {\bibfnamefont {R.~D.}\ \bibnamefont {Desautels}}, \bibinfo {author} {\bibfnamefont {L.}~\bibnamefont {DeBeer-Schmitt}}, \bibinfo {author} {\bibfnamefont {S.~A.}\ \bibnamefont {Montoya}}, \bibinfo {author} {\bibfnamefont {J.~A.}\ \bibnamefont {Borchers}}, \bibinfo {author} {\bibfnamefont {S.-G.}\ \bibnamefont {Je}}, \bibinfo {author} {\bibfnamefont {N.}~\bibnamefont {Tang}}, \bibinfo {author} {\bibfnamefont {M.-Y.}\ \bibnamefont {Im}}, \bibinfo {author} {\bibfnamefont {M.~R.}\ \bibnamefont {Fitzsimmons}}, \bibinfo {author} {\bibfnamefont {E.~E.}\ \bibnamefont {Fullerton}},\ and\ \bibinfo {author} {\bibfnamefont {D.~A.}\ \bibnamefont {Gilbert}},\ }\href@noop {} {\bibfield  {journal} {\bibinfo  {journal} {Physical Review Materials}\ }\textbf {\bibinfo {volume} {3}},\ \bibinfo {pages} {104406} (\bibinfo {year} {2019})}\BibitemShut {NoStop}%
\bibitem [{\citenamefont {Hrabec}\ \emph {et~al.}(2017)\citenamefont {Hrabec}, \citenamefont {Sampaio}, \citenamefont {Belmeguenai}, \citenamefont {Gross}, \citenamefont {Weil}, \citenamefont {Ch{\'e}rif}, \citenamefont {Stashkevich}, \citenamefont {Jacques}, \citenamefont {Thiaville},\ and\ \citenamefont {Rohart}}]{hrabec2017current}%
  \BibitemOpen
  \bibfield  {author} {\bibinfo {author} {\bibfnamefont {A.}~\bibnamefont {Hrabec}}, \bibinfo {author} {\bibfnamefont {J.}~\bibnamefont {Sampaio}}, \bibinfo {author} {\bibfnamefont {M.}~\bibnamefont {Belmeguenai}}, \bibinfo {author} {\bibfnamefont {I.}~\bibnamefont {Gross}}, \bibinfo {author} {\bibfnamefont {R.}~\bibnamefont {Weil}}, \bibinfo {author} {\bibfnamefont {S.~M.}\ \bibnamefont {Ch{\'e}rif}}, \bibinfo {author} {\bibfnamefont {A.}~\bibnamefont {Stashkevich}}, \bibinfo {author} {\bibfnamefont {V.}~\bibnamefont {Jacques}}, \bibinfo {author} {\bibfnamefont {A.}~\bibnamefont {Thiaville}},\ and\ \bibinfo {author} {\bibfnamefont {S.}~\bibnamefont {Rohart}},\ }\href@noop {} {\bibfield  {journal} {\bibinfo  {journal} {Nature communications}\ }\textbf {\bibinfo {volume} {8}},\ \bibinfo {pages} {15765} (\bibinfo {year} {2017})}\BibitemShut {NoStop}%
\bibitem [{\citenamefont {Zhang}\ \emph {et~al.}(2020)\citenamefont {Zhang}, \citenamefont {Zhang}, \citenamefont {Chen}, \citenamefont {Guang}, \citenamefont {Zeng}, \citenamefont {Yu}, \citenamefont {Zhang}, \citenamefont {Liu}, \citenamefont {Feng}, \citenamefont {Zhao} \emph {et~al.}}]{zhang2020formation}%
  \BibitemOpen
  \bibfield  {author} {\bibinfo {author} {\bibfnamefont {J.}~\bibnamefont {Zhang}}, \bibinfo {author} {\bibfnamefont {X.}~\bibnamefont {Zhang}}, \bibinfo {author} {\bibfnamefont {H.}~\bibnamefont {Chen}}, \bibinfo {author} {\bibfnamefont {Y.}~\bibnamefont {Guang}}, \bibinfo {author} {\bibfnamefont {X.}~\bibnamefont {Zeng}}, \bibinfo {author} {\bibfnamefont {G.}~\bibnamefont {Yu}}, \bibinfo {author} {\bibfnamefont {S.}~\bibnamefont {Zhang}}, \bibinfo {author} {\bibfnamefont {Y.}~\bibnamefont {Liu}}, \bibinfo {author} {\bibfnamefont {J.}~\bibnamefont {Feng}}, \bibinfo {author} {\bibfnamefont {Y.}~\bibnamefont {Zhao}}, \emph {et~al.},\ }\href@noop {} {\bibfield  {journal} {\bibinfo  {journal} {Applied Physics Letters}\ }\textbf {\bibinfo {volume} {116}} (\bibinfo {year} {2020})}\BibitemShut {NoStop}%
\bibitem [{\citenamefont {Heigl}\ \emph {et~al.}(2021)\citenamefont {Heigl}, \citenamefont {Koraltan}, \citenamefont {Va{\v{n}}atka}, \citenamefont {Kraft}, \citenamefont {Abert}, \citenamefont {Vogler}, \citenamefont {Semisalova}, \citenamefont {Che}, \citenamefont {Ullrich}, \citenamefont {Schmidt} \emph {et~al.}}]{heigl2021dipolar}%
  \BibitemOpen
  \bibfield  {author} {\bibinfo {author} {\bibfnamefont {M.}~\bibnamefont {Heigl}}, \bibinfo {author} {\bibfnamefont {S.}~\bibnamefont {Koraltan}}, \bibinfo {author} {\bibfnamefont {M.}~\bibnamefont {Va{\v{n}}atka}}, \bibinfo {author} {\bibfnamefont {R.}~\bibnamefont {Kraft}}, \bibinfo {author} {\bibfnamefont {C.}~\bibnamefont {Abert}}, \bibinfo {author} {\bibfnamefont {C.}~\bibnamefont {Vogler}}, \bibinfo {author} {\bibfnamefont {A.}~\bibnamefont {Semisalova}}, \bibinfo {author} {\bibfnamefont {P.}~\bibnamefont {Che}}, \bibinfo {author} {\bibfnamefont {A.}~\bibnamefont {Ullrich}}, \bibinfo {author} {\bibfnamefont {T.}~\bibnamefont {Schmidt}}, \emph {et~al.},\ }\href@noop {} {\bibfield  {journal} {\bibinfo  {journal} {Nature Communications}\ }\textbf {\bibinfo {volume} {12}},\ \bibinfo {pages} {2611} (\bibinfo {year} {2021})}\BibitemShut {NoStop}%
\bibitem [{\citenamefont {Ezawa}(2010)}]{ezawa2010giant}%
  \BibitemOpen
  \bibfield  {author} {\bibinfo {author} {\bibfnamefont {M.}~\bibnamefont {Ezawa}},\ }\href@noop {} {\bibfield  {journal} {\bibinfo  {journal} {Physical review letters}\ }\textbf {\bibinfo {volume} {105}},\ \bibinfo {pages} {197202} (\bibinfo {year} {2010})}\BibitemShut {NoStop}%
\bibitem [{\citenamefont {Dohi}\ \emph {et~al.}(2019)\citenamefont {Dohi}, \citenamefont {DuttaGupta}, \citenamefont {Fukami},\ and\ \citenamefont {Ohno}}]{dohi2019formation}%
  \BibitemOpen
  \bibfield  {author} {\bibinfo {author} {\bibfnamefont {T.}~\bibnamefont {Dohi}}, \bibinfo {author} {\bibfnamefont {S.}~\bibnamefont {DuttaGupta}}, \bibinfo {author} {\bibfnamefont {S.}~\bibnamefont {Fukami}},\ and\ \bibinfo {author} {\bibfnamefont {H.}~\bibnamefont {Ohno}},\ }\href@noop {} {\bibfield  {journal} {\bibinfo  {journal} {Nature communications}\ }\textbf {\bibinfo {volume} {10}},\ \bibinfo {pages} {5153} (\bibinfo {year} {2019})}\BibitemShut {NoStop}%
\bibitem [{\citenamefont {Zazvorka}\ \emph {et~al.}(2020)\citenamefont {Zazvorka}, \citenamefont {Dittrich}, \citenamefont {Ge}, \citenamefont {Kerber}, \citenamefont {Raab}, \citenamefont {Winkler}, \citenamefont {Litzius}, \citenamefont {Veis}, \citenamefont {Virnau},\ and\ \citenamefont {Kl{\"a}ui}}]{zazvorka2020skyrmion}%
  \BibitemOpen
  \bibfield  {author} {\bibinfo {author} {\bibfnamefont {J.}~\bibnamefont {Zazvorka}}, \bibinfo {author} {\bibfnamefont {F.}~\bibnamefont {Dittrich}}, \bibinfo {author} {\bibfnamefont {Y.}~\bibnamefont {Ge}}, \bibinfo {author} {\bibfnamefont {N.}~\bibnamefont {Kerber}}, \bibinfo {author} {\bibfnamefont {K.}~\bibnamefont {Raab}}, \bibinfo {author} {\bibfnamefont {T.}~\bibnamefont {Winkler}}, \bibinfo {author} {\bibfnamefont {K.}~\bibnamefont {Litzius}}, \bibinfo {author} {\bibfnamefont {M.}~\bibnamefont {Veis}}, \bibinfo {author} {\bibfnamefont {P.}~\bibnamefont {Virnau}},\ and\ \bibinfo {author} {\bibfnamefont {M.}~\bibnamefont {Kl{\"a}ui}},\ }\href@noop {} {\bibfield  {journal} {\bibinfo  {journal} {Advanced Functional Materials}\ }\textbf {\bibinfo {volume} {30}},\ \bibinfo {pages} {2004037} (\bibinfo {year} {2020})}\BibitemShut {NoStop}%
\bibitem [{\citenamefont {Gruber}\ \emph {et~al.}(2022)\citenamefont {Gruber}, \citenamefont {Z{\'a}zvorka}, \citenamefont {Brems}, \citenamefont {Rodrigues}, \citenamefont {Dohi}, \citenamefont {Kerber}, \citenamefont {Seng}, \citenamefont {Vafaee}, \citenamefont {Everschor-Sitte}, \citenamefont {Virnau} \emph {et~al.}}]{gruber2022skyrmion}%
  \BibitemOpen
  \bibfield  {author} {\bibinfo {author} {\bibfnamefont {R.}~\bibnamefont {Gruber}}, \bibinfo {author} {\bibfnamefont {J.}~\bibnamefont {Z{\'a}zvorka}}, \bibinfo {author} {\bibfnamefont {M.~A.}\ \bibnamefont {Brems}}, \bibinfo {author} {\bibfnamefont {D.~R.}\ \bibnamefont {Rodrigues}}, \bibinfo {author} {\bibfnamefont {T.}~\bibnamefont {Dohi}}, \bibinfo {author} {\bibfnamefont {N.}~\bibnamefont {Kerber}}, \bibinfo {author} {\bibfnamefont {B.}~\bibnamefont {Seng}}, \bibinfo {author} {\bibfnamefont {M.}~\bibnamefont {Vafaee}}, \bibinfo {author} {\bibfnamefont {K.}~\bibnamefont {Everschor-Sitte}}, \bibinfo {author} {\bibfnamefont {P.}~\bibnamefont {Virnau}}, \emph {et~al.},\ }\href@noop {} {\bibfield  {journal} {\bibinfo  {journal} {Nature Communications}\ }\textbf {\bibinfo {volume} {13}},\ \bibinfo {pages} {3144} (\bibinfo {year} {2022})}\BibitemShut {NoStop}%
\bibitem [{\citenamefont {Kent}\ \emph {et~al.}(2019)\citenamefont {Kent}, \citenamefont {Streubel}, \citenamefont {Lambert}, \citenamefont {Ceballos}, \citenamefont {Je}, \citenamefont {Dhuey}, \citenamefont {Im}, \citenamefont {B{\"u}ttner}, \citenamefont {Hellman}, \citenamefont {Salahuddin} \emph {et~al.}}]{kent2019generation}%
  \BibitemOpen
  \bibfield  {author} {\bibinfo {author} {\bibfnamefont {N.}~\bibnamefont {Kent}}, \bibinfo {author} {\bibfnamefont {R.}~\bibnamefont {Streubel}}, \bibinfo {author} {\bibfnamefont {C.-H.}\ \bibnamefont {Lambert}}, \bibinfo {author} {\bibfnamefont {A.}~\bibnamefont {Ceballos}}, \bibinfo {author} {\bibfnamefont {S.-G.}\ \bibnamefont {Je}}, \bibinfo {author} {\bibfnamefont {S.}~\bibnamefont {Dhuey}}, \bibinfo {author} {\bibfnamefont {M.-Y.}\ \bibnamefont {Im}}, \bibinfo {author} {\bibfnamefont {F.}~\bibnamefont {B{\"u}ttner}}, \bibinfo {author} {\bibfnamefont {F.}~\bibnamefont {Hellman}}, \bibinfo {author} {\bibfnamefont {S.}~\bibnamefont {Salahuddin}}, \emph {et~al.},\ }\href@noop {} {\bibfield  {journal} {\bibinfo  {journal} {Applied Physics Letters}\ }\textbf {\bibinfo {volume} {115}} (\bibinfo {year} {2019})}\BibitemShut {NoStop}%
\bibitem [{\citenamefont {Kent}\ \emph {et~al.}(2021)\citenamefont {Kent}, \citenamefont {Reynolds}, \citenamefont {Raftrey}, \citenamefont {Campbell}, \citenamefont {Virasawmy}, \citenamefont {Dhuey}, \citenamefont {Chopdekar}, \citenamefont {Hierro-Rodriguez}, \citenamefont {Sorrentino}, \citenamefont {Pereiro} \emph {et~al.}}]{kent2021creation}%
  \BibitemOpen
  \bibfield  {author} {\bibinfo {author} {\bibfnamefont {N.}~\bibnamefont {Kent}}, \bibinfo {author} {\bibfnamefont {N.}~\bibnamefont {Reynolds}}, \bibinfo {author} {\bibfnamefont {D.}~\bibnamefont {Raftrey}}, \bibinfo {author} {\bibfnamefont {I.~T.}\ \bibnamefont {Campbell}}, \bibinfo {author} {\bibfnamefont {S.}~\bibnamefont {Virasawmy}}, \bibinfo {author} {\bibfnamefont {S.}~\bibnamefont {Dhuey}}, \bibinfo {author} {\bibfnamefont {R.~V.}\ \bibnamefont {Chopdekar}}, \bibinfo {author} {\bibfnamefont {A.}~\bibnamefont {Hierro-Rodriguez}}, \bibinfo {author} {\bibfnamefont {A.}~\bibnamefont {Sorrentino}}, \bibinfo {author} {\bibfnamefont {E.}~\bibnamefont {Pereiro}}, \emph {et~al.},\ }\href@noop {} {\bibfield  {journal} {\bibinfo  {journal} {Nature communications}\ }\textbf {\bibinfo {volume} {12}},\ \bibinfo {pages} {1562} (\bibinfo {year} {2021})}\BibitemShut {NoStop}%
\bibitem [{\citenamefont {Grelier}\ \emph {et~al.}(2022)\citenamefont {Grelier}, \citenamefont {Godel}, \citenamefont {Vecchiola}, \citenamefont {Collin}, \citenamefont {Bouzehouane}, \citenamefont {Fert}, \citenamefont {Cros},\ and\ \citenamefont {Reyren}}]{grelier2022three}%
  \BibitemOpen
  \bibfield  {author} {\bibinfo {author} {\bibfnamefont {M.}~\bibnamefont {Grelier}}, \bibinfo {author} {\bibfnamefont {F.}~\bibnamefont {Godel}}, \bibinfo {author} {\bibfnamefont {A.}~\bibnamefont {Vecchiola}}, \bibinfo {author} {\bibfnamefont {S.}~\bibnamefont {Collin}}, \bibinfo {author} {\bibfnamefont {K.}~\bibnamefont {Bouzehouane}}, \bibinfo {author} {\bibfnamefont {A.}~\bibnamefont {Fert}}, \bibinfo {author} {\bibfnamefont {V.}~\bibnamefont {Cros}},\ and\ \bibinfo {author} {\bibfnamefont {N.}~\bibnamefont {Reyren}},\ }\href@noop {} {\bibfield  {journal} {\bibinfo  {journal} {Nature communications}\ }\textbf {\bibinfo {volume} {13}},\ \bibinfo {pages} {6843} (\bibinfo {year} {2022})}\BibitemShut {NoStop}%
\bibitem [{\citenamefont {Yu}\ \emph {et~al.}(2024)\citenamefont {Yu}, \citenamefont {Nakanishi}, \citenamefont {Chiew}, \citenamefont {Liu}, \citenamefont {Nakajima}, \citenamefont {Kanazawa}, \citenamefont {Karube}, \citenamefont {Taguchi}, \citenamefont {Nagaosa},\ and\ \citenamefont {Tokura}}]{yu20243d}%
  \BibitemOpen
  \bibfield  {author} {\bibinfo {author} {\bibfnamefont {X.}~\bibnamefont {Yu}}, \bibinfo {author} {\bibfnamefont {N.}~\bibnamefont {Nakanishi}}, \bibinfo {author} {\bibfnamefont {Y.-L.}\ \bibnamefont {Chiew}}, \bibinfo {author} {\bibfnamefont {Y.}~\bibnamefont {Liu}}, \bibinfo {author} {\bibfnamefont {K.}~\bibnamefont {Nakajima}}, \bibinfo {author} {\bibfnamefont {N.}~\bibnamefont {Kanazawa}}, \bibinfo {author} {\bibfnamefont {K.}~\bibnamefont {Karube}}, \bibinfo {author} {\bibfnamefont {Y.}~\bibnamefont {Taguchi}}, \bibinfo {author} {\bibfnamefont {N.}~\bibnamefont {Nagaosa}},\ and\ \bibinfo {author} {\bibfnamefont {Y.}~\bibnamefont {Tokura}},\ }\href@noop {} {\bibfield  {journal} {\bibinfo  {journal} {Communications Materials}\ }\textbf {\bibinfo {volume} {5}},\ \bibinfo {pages} {80} (\bibinfo {year} {2024})}\BibitemShut {NoStop}%
\bibitem [{\citenamefont {Wang}\ \emph {et~al.}(2018)\citenamefont {Wang}, \citenamefont {Yuan},\ and\ \citenamefont {Wang}}]{wang2018theory}%
  \BibitemOpen
  \bibfield  {author} {\bibinfo {author} {\bibfnamefont {X.}~\bibnamefont {Wang}}, \bibinfo {author} {\bibfnamefont {H.}~\bibnamefont {Yuan}},\ and\ \bibinfo {author} {\bibfnamefont {X.}~\bibnamefont {Wang}},\ }\href@noop {} {\bibfield  {journal} {\bibinfo  {journal} {Communications Physics}\ }\textbf {\bibinfo {volume} {1}},\ \bibinfo {pages} {31} (\bibinfo {year} {2018})}\BibitemShut {NoStop}%
\bibitem [{\citenamefont {Rohart}\ and\ \citenamefont {Thiaville}(2013)}]{rohart2013skyrmion}%
  \BibitemOpen
  \bibfield  {author} {\bibinfo {author} {\bibfnamefont {S.}~\bibnamefont {Rohart}}\ and\ \bibinfo {author} {\bibfnamefont {A.}~\bibnamefont {Thiaville}},\ }\href@noop {} {\bibfield  {journal} {\bibinfo  {journal} {Physical Review B}\ }\textbf {\bibinfo {volume} {88}},\ \bibinfo {pages} {184422} (\bibinfo {year} {2013})}\BibitemShut {NoStop}%
\bibitem [{\citenamefont {Li}\ \emph {et~al.}(2019)\citenamefont {Li}, \citenamefont {Bykova}, \citenamefont {Zhang}, \citenamefont {Yu}, \citenamefont {Tomasello}, \citenamefont {Carpentieri}, \citenamefont {Liu}, \citenamefont {Guang}, \citenamefont {Gr{\"a}fe}, \citenamefont {Weigand} \emph {et~al.}}]{li2019anatomy}%
  \BibitemOpen
  \bibfield  {author} {\bibinfo {author} {\bibfnamefont {W.}~\bibnamefont {Li}}, \bibinfo {author} {\bibfnamefont {I.}~\bibnamefont {Bykova}}, \bibinfo {author} {\bibfnamefont {S.}~\bibnamefont {Zhang}}, \bibinfo {author} {\bibfnamefont {G.}~\bibnamefont {Yu}}, \bibinfo {author} {\bibfnamefont {R.}~\bibnamefont {Tomasello}}, \bibinfo {author} {\bibfnamefont {M.}~\bibnamefont {Carpentieri}}, \bibinfo {author} {\bibfnamefont {Y.}~\bibnamefont {Liu}}, \bibinfo {author} {\bibfnamefont {Y.}~\bibnamefont {Guang}}, \bibinfo {author} {\bibfnamefont {J.}~\bibnamefont {Gr{\"a}fe}}, \bibinfo {author} {\bibfnamefont {M.}~\bibnamefont {Weigand}}, \emph {et~al.},\ }\href@noop {} {\bibfield  {journal} {\bibinfo  {journal} {Advanced Materials}\ }\textbf {\bibinfo {volume} {31}},\ \bibinfo {pages} {1807683} (\bibinfo {year} {2019})}\BibitemShut {NoStop}%
\bibitem [{\citenamefont {Vidal-Silva}\ \emph {et~al.}(2017)\citenamefont {Vidal-Silva}, \citenamefont {Riveros},\ and\ \citenamefont {Escrig}}]{vidal2017stability}%
  \BibitemOpen
  \bibfield  {author} {\bibinfo {author} {\bibfnamefont {N.}~\bibnamefont {Vidal-Silva}}, \bibinfo {author} {\bibfnamefont {A.}~\bibnamefont {Riveros}},\ and\ \bibinfo {author} {\bibfnamefont {J.}~\bibnamefont {Escrig}},\ }\href@noop {} {\bibfield  {journal} {\bibinfo  {journal} {Journal of Magnetism and Magnetic Materials}\ }\textbf {\bibinfo {volume} {443}},\ \bibinfo {pages} {116} (\bibinfo {year} {2017})}\BibitemShut {NoStop}%
\bibitem [{\citenamefont {Leonov}\ \emph {et~al.}(2016)\citenamefont {Leonov}, \citenamefont {Monchesky}, \citenamefont {Romming}, \citenamefont {Kubetzka}, \citenamefont {Bogdanov},\ and\ \citenamefont {Wiesendanger}}]{leonov2016properties}%
  \BibitemOpen
  \bibfield  {author} {\bibinfo {author} {\bibfnamefont {A.}~\bibnamefont {Leonov}}, \bibinfo {author} {\bibfnamefont {T.}~\bibnamefont {Monchesky}}, \bibinfo {author} {\bibfnamefont {N.}~\bibnamefont {Romming}}, \bibinfo {author} {\bibfnamefont {A.}~\bibnamefont {Kubetzka}}, \bibinfo {author} {\bibfnamefont {A.}~\bibnamefont {Bogdanov}},\ and\ \bibinfo {author} {\bibfnamefont {R.}~\bibnamefont {Wiesendanger}},\ }\href@noop {} {\bibfield  {journal} {\bibinfo  {journal} {New Journal of Physics}\ }\textbf {\bibinfo {volume} {18}},\ \bibinfo {pages} {065003} (\bibinfo {year} {2016})}\BibitemShut {NoStop}%
\bibitem [{\citenamefont {B{\"u}ttner}\ \emph {et~al.}(2018)\citenamefont {B{\"u}ttner}, \citenamefont {Lemesh},\ and\ \citenamefont {Beach}}]{buttner2018theory}%
  \BibitemOpen
  \bibfield  {author} {\bibinfo {author} {\bibfnamefont {F.}~\bibnamefont {B{\"u}ttner}}, \bibinfo {author} {\bibfnamefont {I.}~\bibnamefont {Lemesh}},\ and\ \bibinfo {author} {\bibfnamefont {G.~S.}\ \bibnamefont {Beach}},\ }\href@noop {} {\bibfield  {journal} {\bibinfo  {journal} {Scientific reports}\ }\textbf {\bibinfo {volume} {8}},\ \bibinfo {pages} {4464} (\bibinfo {year} {2018})}\BibitemShut {NoStop}%
\bibitem [{\citenamefont {Prychynenko}\ \emph {et~al.}(2018)\citenamefont {Prychynenko}, \citenamefont {Sitte}, \citenamefont {Litzius}, \citenamefont {Kr{\"u}ger}, \citenamefont {Bourianoff}, \citenamefont {Kl{\"a}ui}, \citenamefont {Sinova},\ and\ \citenamefont {Everschor-Sitte}}]{prychynenko2018magnetic}%
  \BibitemOpen
  \bibfield  {author} {\bibinfo {author} {\bibfnamefont {D.}~\bibnamefont {Prychynenko}}, \bibinfo {author} {\bibfnamefont {M.}~\bibnamefont {Sitte}}, \bibinfo {author} {\bibfnamefont {K.}~\bibnamefont {Litzius}}, \bibinfo {author} {\bibfnamefont {B.}~\bibnamefont {Kr{\"u}ger}}, \bibinfo {author} {\bibfnamefont {G.}~\bibnamefont {Bourianoff}}, \bibinfo {author} {\bibfnamefont {M.}~\bibnamefont {Kl{\"a}ui}}, \bibinfo {author} {\bibfnamefont {J.}~\bibnamefont {Sinova}},\ and\ \bibinfo {author} {\bibfnamefont {K.}~\bibnamefont {Everschor-Sitte}},\ }\href@noop {} {\bibfield  {journal} {\bibinfo  {journal} {Physical Review Applied}\ }\textbf {\bibinfo {volume} {9}},\ \bibinfo {pages} {014034} (\bibinfo {year} {2018})}\BibitemShut {NoStop}%
\bibitem [{\citenamefont {Sun}\ \emph {et~al.}(2023)\citenamefont {Sun}, \citenamefont {Lin}, \citenamefont {Lei}, \citenamefont {Chen}, \citenamefont {Kang}, \citenamefont {Zhao}, \citenamefont {Wei}, \citenamefont {Chen}, \citenamefont {Pang}, \citenamefont {Hu} \emph {et~al.}}]{sun2023experimental}%
  \BibitemOpen
  \bibfield  {author} {\bibinfo {author} {\bibfnamefont {Y.}~\bibnamefont {Sun}}, \bibinfo {author} {\bibfnamefont {T.}~\bibnamefont {Lin}}, \bibinfo {author} {\bibfnamefont {N.}~\bibnamefont {Lei}}, \bibinfo {author} {\bibfnamefont {X.}~\bibnamefont {Chen}}, \bibinfo {author} {\bibfnamefont {W.}~\bibnamefont {Kang}}, \bibinfo {author} {\bibfnamefont {Z.}~\bibnamefont {Zhao}}, \bibinfo {author} {\bibfnamefont {D.}~\bibnamefont {Wei}}, \bibinfo {author} {\bibfnamefont {C.}~\bibnamefont {Chen}}, \bibinfo {author} {\bibfnamefont {S.}~\bibnamefont {Pang}}, \bibinfo {author} {\bibfnamefont {L.}~\bibnamefont {Hu}}, \emph {et~al.},\ }\href@noop {} {\bibfield  {journal} {\bibinfo  {journal} {Nature Communications}\ }\textbf {\bibinfo {volume} {14}},\ \bibinfo {pages} {3434} (\bibinfo {year} {2023})}\BibitemShut {NoStop}%
\bibitem [{\citenamefont {Lemesh}\ \emph {et~al.}(2017)\citenamefont {Lemesh}, \citenamefont {B{\"u}ttner},\ and\ \citenamefont {Beach}}]{lemesh2017accurate}%
  \BibitemOpen
  \bibfield  {author} {\bibinfo {author} {\bibfnamefont {I.}~\bibnamefont {Lemesh}}, \bibinfo {author} {\bibfnamefont {F.}~\bibnamefont {B{\"u}ttner}},\ and\ \bibinfo {author} {\bibfnamefont {G.~S.}\ \bibnamefont {Beach}},\ }\href@noop {} {\bibfield  {journal} {\bibinfo  {journal} {Physical Review B}\ }\textbf {\bibinfo {volume} {95}},\ \bibinfo {pages} {174423} (\bibinfo {year} {2017})}\BibitemShut {NoStop}%
\bibitem [{\citenamefont {Cape}\ and\ \citenamefont {Lehman}(1971)}]{cape1971magnetic}%
  \BibitemOpen
  \bibfield  {author} {\bibinfo {author} {\bibfnamefont {J.}~\bibnamefont {Cape}}\ and\ \bibinfo {author} {\bibfnamefont {G.}~\bibnamefont {Lehman}},\ }\href@noop {} {\bibfield  {journal} {\bibinfo  {journal} {Journal of Applied Physics}\ }\textbf {\bibinfo {volume} {42}},\ \bibinfo {pages} {5732} (\bibinfo {year} {1971})}\BibitemShut {NoStop}%
\bibitem [{\citenamefont {Kaplan}\ and\ \citenamefont {Gehring}(1993)}]{kaplan1993domain}%
  \BibitemOpen
  \bibfield  {author} {\bibinfo {author} {\bibfnamefont {B.}~\bibnamefont {Kaplan}}\ and\ \bibinfo {author} {\bibfnamefont {G.}~\bibnamefont {Gehring}},\ }\href@noop {} {\bibfield  {journal} {\bibinfo  {journal} {Journal of Magnetism and Magnetic Materials}\ }\textbf {\bibinfo {volume} {128}},\ \bibinfo {pages} {111} (\bibinfo {year} {1993})}\BibitemShut {NoStop}%
\bibitem [{\citenamefont {Lemesh}\ \emph {et~al.}(2018)\citenamefont {Lemesh}, \citenamefont {Litzius}, \citenamefont {B{\"o}ttcher}, \citenamefont {Bassirian}, \citenamefont {Kerber}, \citenamefont {Heinze}, \citenamefont {Z{\'a}zvorka}, \citenamefont {B{\"u}ttner}, \citenamefont {Caretta}, \citenamefont {Mann} \emph {et~al.}}]{lemesh2018current}%
  \BibitemOpen
  \bibfield  {author} {\bibinfo {author} {\bibfnamefont {I.}~\bibnamefont {Lemesh}}, \bibinfo {author} {\bibfnamefont {K.}~\bibnamefont {Litzius}}, \bibinfo {author} {\bibfnamefont {M.}~\bibnamefont {B{\"o}ttcher}}, \bibinfo {author} {\bibfnamefont {P.}~\bibnamefont {Bassirian}}, \bibinfo {author} {\bibfnamefont {N.}~\bibnamefont {Kerber}}, \bibinfo {author} {\bibfnamefont {D.}~\bibnamefont {Heinze}}, \bibinfo {author} {\bibfnamefont {J.}~\bibnamefont {Z{\'a}zvorka}}, \bibinfo {author} {\bibfnamefont {F.}~\bibnamefont {B{\"u}ttner}}, \bibinfo {author} {\bibfnamefont {L.}~\bibnamefont {Caretta}}, \bibinfo {author} {\bibfnamefont {M.}~\bibnamefont {Mann}}, \emph {et~al.},\ }\href@noop {} {\bibfield  {journal} {\bibinfo  {journal} {Advanced Materials}\ }\textbf {\bibinfo {volume} {30}},\ \bibinfo {pages} {1805461} (\bibinfo {year} {2018})}\BibitemShut {NoStop}%
\bibitem [{\citenamefont {Zazvorka}\ \emph {et~al.}(2019)\citenamefont {Zazvorka}, \citenamefont {Jakobs}, \citenamefont {Heinze}, \citenamefont {Keil}, \citenamefont {Kromin}, \citenamefont {Jaiswal}, \citenamefont {Litzius}, \citenamefont {Jakob}, \citenamefont {Virnau}, \citenamefont {Pinna} \emph {et~al.}}]{zazvorka2019thermal}%
  \BibitemOpen
  \bibfield  {author} {\bibinfo {author} {\bibfnamefont {J.}~\bibnamefont {Zazvorka}}, \bibinfo {author} {\bibfnamefont {F.}~\bibnamefont {Jakobs}}, \bibinfo {author} {\bibfnamefont {D.}~\bibnamefont {Heinze}}, \bibinfo {author} {\bibfnamefont {N.}~\bibnamefont {Keil}}, \bibinfo {author} {\bibfnamefont {S.}~\bibnamefont {Kromin}}, \bibinfo {author} {\bibfnamefont {S.}~\bibnamefont {Jaiswal}}, \bibinfo {author} {\bibfnamefont {K.}~\bibnamefont {Litzius}}, \bibinfo {author} {\bibfnamefont {G.}~\bibnamefont {Jakob}}, \bibinfo {author} {\bibfnamefont {P.}~\bibnamefont {Virnau}}, \bibinfo {author} {\bibfnamefont {D.}~\bibnamefont {Pinna}}, \emph {et~al.},\ }\href@noop {} {\bibfield  {journal} {\bibinfo  {journal} {Nature nanotechnology}\ }\textbf {\bibinfo {volume} {14}},\ \bibinfo {pages} {658} (\bibinfo {year} {2019})}\BibitemShut {NoStop}%
\bibitem [{\citenamefont {Ge}\ \emph {et~al.}(2023)\citenamefont {Ge}, \citenamefont {Roth{\"o}rl}, \citenamefont {Brems}, \citenamefont {Kerber}, \citenamefont {Gruber}, \citenamefont {Dohi}, \citenamefont {Kl{\"a}ui},\ and\ \citenamefont {Virnau}}]{ge2023constructing}%
  \BibitemOpen
  \bibfield  {author} {\bibinfo {author} {\bibfnamefont {Y.}~\bibnamefont {Ge}}, \bibinfo {author} {\bibfnamefont {J.}~\bibnamefont {Roth{\"o}rl}}, \bibinfo {author} {\bibfnamefont {M.~A.}\ \bibnamefont {Brems}}, \bibinfo {author} {\bibfnamefont {N.}~\bibnamefont {Kerber}}, \bibinfo {author} {\bibfnamefont {R.}~\bibnamefont {Gruber}}, \bibinfo {author} {\bibfnamefont {T.}~\bibnamefont {Dohi}}, \bibinfo {author} {\bibfnamefont {M.}~\bibnamefont {Kl{\"a}ui}},\ and\ \bibinfo {author} {\bibfnamefont {P.}~\bibnamefont {Virnau}},\ }\href@noop {} {\bibfield  {journal} {\bibinfo  {journal} {Communications Physics}\ }\textbf {\bibinfo {volume} {6}},\ \bibinfo {pages} {30} (\bibinfo {year} {2023})}\BibitemShut {NoStop}%
\bibitem [{\citenamefont {Woo}\ \emph {et~al.}(2016)\citenamefont {Woo}, \citenamefont {Litzius}, \citenamefont {Kr{\"u}ger}, \citenamefont {Im}, \citenamefont {Caretta}, \citenamefont {Richter}, \citenamefont {Mann}, \citenamefont {Krone}, \citenamefont {Reeve}, \citenamefont {Weigand} \emph {et~al.}}]{woo2016observation}%
  \BibitemOpen
  \bibfield  {author} {\bibinfo {author} {\bibfnamefont {S.}~\bibnamefont {Woo}}, \bibinfo {author} {\bibfnamefont {K.}~\bibnamefont {Litzius}}, \bibinfo {author} {\bibfnamefont {B.}~\bibnamefont {Kr{\"u}ger}}, \bibinfo {author} {\bibfnamefont {M.-Y.}\ \bibnamefont {Im}}, \bibinfo {author} {\bibfnamefont {L.}~\bibnamefont {Caretta}}, \bibinfo {author} {\bibfnamefont {K.}~\bibnamefont {Richter}}, \bibinfo {author} {\bibfnamefont {M.}~\bibnamefont {Mann}}, \bibinfo {author} {\bibfnamefont {A.}~\bibnamefont {Krone}}, \bibinfo {author} {\bibfnamefont {R.~M.}\ \bibnamefont {Reeve}}, \bibinfo {author} {\bibfnamefont {M.}~\bibnamefont {Weigand}}, \emph {et~al.},\ }\href@noop {} {\bibfield  {journal} {\bibinfo  {journal} {Nature materials}\ }\textbf {\bibinfo {volume} {15}},\ \bibinfo {pages} {501} (\bibinfo {year} {2016})}\BibitemShut {NoStop}%
\bibitem [{\citenamefont {B{\"u}ttner}\ \emph {et~al.}(2017)\citenamefont {B{\"u}ttner}, \citenamefont {Lemesh}, \citenamefont {Schneider}, \citenamefont {Pfau}, \citenamefont {G{\"u}nther}, \citenamefont {Hessing}, \citenamefont {Geilhufe}, \citenamefont {Caretta}, \citenamefont {Engel}, \citenamefont {Kr{\"u}ger} \emph {et~al.}}]{buttner2017field}%
  \BibitemOpen
  \bibfield  {author} {\bibinfo {author} {\bibfnamefont {F.}~\bibnamefont {B{\"u}ttner}}, \bibinfo {author} {\bibfnamefont {I.}~\bibnamefont {Lemesh}}, \bibinfo {author} {\bibfnamefont {M.}~\bibnamefont {Schneider}}, \bibinfo {author} {\bibfnamefont {B.}~\bibnamefont {Pfau}}, \bibinfo {author} {\bibfnamefont {C.~M.}\ \bibnamefont {G{\"u}nther}}, \bibinfo {author} {\bibfnamefont {P.}~\bibnamefont {Hessing}}, \bibinfo {author} {\bibfnamefont {J.}~\bibnamefont {Geilhufe}}, \bibinfo {author} {\bibfnamefont {L.}~\bibnamefont {Caretta}}, \bibinfo {author} {\bibfnamefont {D.}~\bibnamefont {Engel}}, \bibinfo {author} {\bibfnamefont {B.}~\bibnamefont {Kr{\"u}ger}}, \emph {et~al.},\ }\href@noop {} {\bibfield  {journal} {\bibinfo  {journal} {Nature Nanotechnology}\ }\textbf {\bibinfo {volume} {12}},\ \bibinfo {pages} {1040} (\bibinfo {year} {2017})}\BibitemShut {NoStop}%
\bibitem [{\citenamefont {Romming}\ \emph {et~al.}(2015)\citenamefont {Romming}, \citenamefont {Kubetzka}, \citenamefont {Hanneken}, \citenamefont {von Bergmann},\ and\ \citenamefont {Wiesendanger}}]{romming2015field}%
  \BibitemOpen
  \bibfield  {author} {\bibinfo {author} {\bibfnamefont {N.}~\bibnamefont {Romming}}, \bibinfo {author} {\bibfnamefont {A.}~\bibnamefont {Kubetzka}}, \bibinfo {author} {\bibfnamefont {C.}~\bibnamefont {Hanneken}}, \bibinfo {author} {\bibfnamefont {K.}~\bibnamefont {von Bergmann}},\ and\ \bibinfo {author} {\bibfnamefont {R.}~\bibnamefont {Wiesendanger}},\ }\href@noop {} {\bibfield  {journal} {\bibinfo  {journal} {Physical review letters}\ }\textbf {\bibinfo {volume} {114}},\ \bibinfo {pages} {177203} (\bibinfo {year} {2015})}\BibitemShut {NoStop}%
\bibitem [{\citenamefont {Boulle}\ \emph {et~al.}(2016)\citenamefont {Boulle}, \citenamefont {Vogel}, \citenamefont {Yang}, \citenamefont {Pizzini}, \citenamefont {de~Souza~Chaves}, \citenamefont {Locatelli}, \citenamefont {Mente{\c{s}}}, \citenamefont {Sala}, \citenamefont {Buda-Prejbeanu}, \citenamefont {Klein} \emph {et~al.}}]{boulle2016room}%
  \BibitemOpen
  \bibfield  {author} {\bibinfo {author} {\bibfnamefont {O.}~\bibnamefont {Boulle}}, \bibinfo {author} {\bibfnamefont {J.}~\bibnamefont {Vogel}}, \bibinfo {author} {\bibfnamefont {H.}~\bibnamefont {Yang}}, \bibinfo {author} {\bibfnamefont {S.}~\bibnamefont {Pizzini}}, \bibinfo {author} {\bibfnamefont {D.}~\bibnamefont {de~Souza~Chaves}}, \bibinfo {author} {\bibfnamefont {A.}~\bibnamefont {Locatelli}}, \bibinfo {author} {\bibfnamefont {T.~O.}\ \bibnamefont {Mente{\c{s}}}}, \bibinfo {author} {\bibfnamefont {A.}~\bibnamefont {Sala}}, \bibinfo {author} {\bibfnamefont {L.~D.}\ \bibnamefont {Buda-Prejbeanu}}, \bibinfo {author} {\bibfnamefont {O.}~\bibnamefont {Klein}}, \emph {et~al.},\ }\href@noop {} {\bibfield  {journal} {\bibinfo  {journal} {Nature nanotechnology}\ }\textbf {\bibinfo {volume} {11}},\ \bibinfo {pages} {449} (\bibinfo {year} {2016})}\BibitemShut {NoStop}%
\bibitem [{\citenamefont {Kac}\ \emph {et~al.}(2002)\citenamefont {Kac}, \citenamefont {Cheung}, \citenamefont {Kac},\ and\ \citenamefont {Cheung}}]{kac2002euler}%
  \BibitemOpen
  \bibfield  {author} {\bibinfo {author} {\bibfnamefont {V.}~\bibnamefont {Kac}}, \bibinfo {author} {\bibfnamefont {P.}~\bibnamefont {Cheung}}, \bibinfo {author} {\bibfnamefont {V.}~\bibnamefont {Kac}},\ and\ \bibinfo {author} {\bibfnamefont {P.}~\bibnamefont {Cheung}},\ }\href@noop {} {\bibfield  {journal} {\bibinfo  {journal} {Quantum Calculus}\ ,\ \bibinfo {pages} {92}} (\bibinfo {year} {2002})}\BibitemShut {NoStop}%
\bibitem [{\citenamefont {Kittel}(1946)}]{kittel1946theory}%
  \BibitemOpen
  \bibfield  {author} {\bibinfo {author} {\bibfnamefont {C.}~\bibnamefont {Kittel}},\ }\href@noop {} {\bibfield  {journal} {\bibinfo  {journal} {Physical Review}\ }\textbf {\bibinfo {volume} {70}},\ \bibinfo {pages} {965} (\bibinfo {year} {1946})}\BibitemShut {NoStop}%
\bibitem [{\citenamefont {Srivastava}\ \emph {et~al.}(2023)\citenamefont {Srivastava}, \citenamefont {Sassi}, \citenamefont {Ajejas}, \citenamefont {Vecchiola}, \citenamefont {Ngouagnia~Yemeli}, \citenamefont {Hurdequint}, \citenamefont {Bouzehouane}, \citenamefont {Reyren}, \citenamefont {Cros}, \citenamefont {Devolder} \emph {et~al.}}]{srivastava2023resonant}%
  \BibitemOpen
  \bibfield  {author} {\bibinfo {author} {\bibfnamefont {T.}~\bibnamefont {Srivastava}}, \bibinfo {author} {\bibfnamefont {Y.}~\bibnamefont {Sassi}}, \bibinfo {author} {\bibfnamefont {F.}~\bibnamefont {Ajejas}}, \bibinfo {author} {\bibfnamefont {A.}~\bibnamefont {Vecchiola}}, \bibinfo {author} {\bibfnamefont {I.}~\bibnamefont {Ngouagnia~Yemeli}}, \bibinfo {author} {\bibfnamefont {H.}~\bibnamefont {Hurdequint}}, \bibinfo {author} {\bibfnamefont {K.}~\bibnamefont {Bouzehouane}}, \bibinfo {author} {\bibfnamefont {N.}~\bibnamefont {Reyren}}, \bibinfo {author} {\bibfnamefont {V.}~\bibnamefont {Cros}}, \bibinfo {author} {\bibfnamefont {T.}~\bibnamefont {Devolder}}, \emph {et~al.},\ }\href@noop {} {\bibfield  {journal} {\bibinfo  {journal} {APL Materials}\ }\textbf {\bibinfo {volume} {11}} (\bibinfo {year} {2023})}\BibitemShut {NoStop}%
\bibitem [{\citenamefont {Liyanage}\ \emph {et~al.}(2023)\citenamefont {Liyanage}, \citenamefont {Tang}, \citenamefont {Quigley}, \citenamefont {Borchers}, \citenamefont {Grutter}, \citenamefont {Maranville}, \citenamefont {Sinha}, \citenamefont {Reyren}, \citenamefont {Montoya}, \citenamefont {Fullerton} \emph {et~al.}}]{liyanage2023three}%
  \BibitemOpen
  \bibfield  {author} {\bibinfo {author} {\bibfnamefont {W.}~\bibnamefont {Liyanage}}, \bibinfo {author} {\bibfnamefont {N.}~\bibnamefont {Tang}}, \bibinfo {author} {\bibfnamefont {L.}~\bibnamefont {Quigley}}, \bibinfo {author} {\bibfnamefont {J.~A.}\ \bibnamefont {Borchers}}, \bibinfo {author} {\bibfnamefont {A.~J.}\ \bibnamefont {Grutter}}, \bibinfo {author} {\bibfnamefont {B.~B.}\ \bibnamefont {Maranville}}, \bibinfo {author} {\bibfnamefont {S.~K.}\ \bibnamefont {Sinha}}, \bibinfo {author} {\bibfnamefont {N.}~\bibnamefont {Reyren}}, \bibinfo {author} {\bibfnamefont {S.~A.}\ \bibnamefont {Montoya}}, \bibinfo {author} {\bibfnamefont {E.~E.}\ \bibnamefont {Fullerton}}, \emph {et~al.},\ }\href@noop {} {\bibfield  {journal} {\bibinfo  {journal} {Physical Review B}\ }\textbf {\bibinfo {volume} {107}},\ \bibinfo {pages} {184412} (\bibinfo {year} {2023})}\BibitemShut {NoStop}%
\bibitem [{\citenamefont {Hubert}\ and\ \citenamefont {Sch{\"a}fer}(1998)}]{hubert1998magnetic}%
  \BibitemOpen
  \bibfield  {author} {\bibinfo {author} {\bibfnamefont {A.}~\bibnamefont {Hubert}}\ and\ \bibinfo {author} {\bibfnamefont {R.}~\bibnamefont {Sch{\"a}fer}},\ }\href@noop {} {\emph {\bibinfo {title} {Magnetic domains: the analysis of magnetic microstructures}}}\ (\bibinfo  {publisher} {Springer Science \& Business Media},\ \bibinfo {year} {1998})\BibitemShut {NoStop}%
\bibitem [{\citenamefont {Sisodia}\ \emph {et~al.}(2021)\citenamefont {Sisodia}, \citenamefont {Muduli}, \citenamefont {Papanicolaou},\ and\ \citenamefont {Komineas}}]{sisodia2021chiral}%
  \BibitemOpen
  \bibfield  {author} {\bibinfo {author} {\bibfnamefont {N.}~\bibnamefont {Sisodia}}, \bibinfo {author} {\bibfnamefont {P.~K.}\ \bibnamefont {Muduli}}, \bibinfo {author} {\bibfnamefont {N.}~\bibnamefont {Papanicolaou}},\ and\ \bibinfo {author} {\bibfnamefont {S.}~\bibnamefont {Komineas}},\ }\href@noop {} {\bibfield  {journal} {\bibinfo  {journal} {Physical Review B}\ }\textbf {\bibinfo {volume} {103}},\ \bibinfo {pages} {024431} (\bibinfo {year} {2021})}\BibitemShut {NoStop}%
\bibitem [{\citenamefont {Pinna}\ \emph {et~al.}(2020)\citenamefont {Pinna}, \citenamefont {Bourianoff},\ and\ \citenamefont {Everschor-Sitte}}]{pinna2020reservoir}%
  \BibitemOpen
  \bibfield  {author} {\bibinfo {author} {\bibfnamefont {D.}~\bibnamefont {Pinna}}, \bibinfo {author} {\bibfnamefont {G.}~\bibnamefont {Bourianoff}},\ and\ \bibinfo {author} {\bibfnamefont {K.}~\bibnamefont {Everschor-Sitte}},\ }\href@noop {} {\bibfield  {journal} {\bibinfo  {journal} {Physical Review Applied}\ }\textbf {\bibinfo {volume} {14}},\ \bibinfo {pages} {054020} (\bibinfo {year} {2020})}\BibitemShut {NoStop}%
\bibitem [{\citenamefont {Iwasaki}\ \emph {et~al.}(2013{\natexlab{b}})\citenamefont {Iwasaki}, \citenamefont {Mochizuki},\ and\ \citenamefont {Nagaosa}}]{iwasaki2013current}%
  \BibitemOpen
  \bibfield  {author} {\bibinfo {author} {\bibfnamefont {J.}~\bibnamefont {Iwasaki}}, \bibinfo {author} {\bibfnamefont {M.}~\bibnamefont {Mochizuki}},\ and\ \bibinfo {author} {\bibfnamefont {N.}~\bibnamefont {Nagaosa}},\ }\href@noop {} {\bibfield  {journal} {\bibinfo  {journal} {Nature nanotechnology}\ }\textbf {\bibinfo {volume} {8}},\ \bibinfo {pages} {742} (\bibinfo {year} {2013}{\natexlab{b}})}\BibitemShut {NoStop}%
\bibitem [{\citenamefont {Litzius}\ \emph {et~al.}(2020)\citenamefont {Litzius}, \citenamefont {Leliaert}, \citenamefont {Bassirian}, \citenamefont {Rodrigues}, \citenamefont {Kromin}, \citenamefont {Lemesh}, \citenamefont {Zazvorka}, \citenamefont {Lee}, \citenamefont {Mulkers}, \citenamefont {Kerber} \emph {et~al.}}]{litzius2020role}%
  \BibitemOpen
  \bibfield  {author} {\bibinfo {author} {\bibfnamefont {K.}~\bibnamefont {Litzius}}, \bibinfo {author} {\bibfnamefont {J.}~\bibnamefont {Leliaert}}, \bibinfo {author} {\bibfnamefont {P.}~\bibnamefont {Bassirian}}, \bibinfo {author} {\bibfnamefont {D.}~\bibnamefont {Rodrigues}}, \bibinfo {author} {\bibfnamefont {S.}~\bibnamefont {Kromin}}, \bibinfo {author} {\bibfnamefont {I.}~\bibnamefont {Lemesh}}, \bibinfo {author} {\bibfnamefont {J.}~\bibnamefont {Zazvorka}}, \bibinfo {author} {\bibfnamefont {K.-J.}\ \bibnamefont {Lee}}, \bibinfo {author} {\bibfnamefont {J.}~\bibnamefont {Mulkers}}, \bibinfo {author} {\bibfnamefont {N.}~\bibnamefont {Kerber}}, \emph {et~al.},\ }\href@noop {} {\bibfield  {journal} {\bibinfo  {journal} {Nature Electronics}\ }\textbf {\bibinfo {volume} {3}},\ \bibinfo {pages} {30} (\bibinfo {year} {2020})}\BibitemShut {NoStop}%
\bibitem [{\citenamefont {Kim}\ and\ \citenamefont {Yoo}(2017)}]{kim2017current}%
  \BibitemOpen
  \bibfield  {author} {\bibinfo {author} {\bibfnamefont {J.-V.}\ \bibnamefont {Kim}}\ and\ \bibinfo {author} {\bibfnamefont {M.-W.}\ \bibnamefont {Yoo}},\ }\href@noop {} {\bibfield  {journal} {\bibinfo  {journal} {Applied Physics Letters}\ }\textbf {\bibinfo {volume} {110}} (\bibinfo {year} {2017})}\BibitemShut {NoStop}%
\bibitem [{\citenamefont {Yu}\ \emph {et~al.}(2016)\citenamefont {Yu}, \citenamefont {Upadhyaya}, \citenamefont {Li}, \citenamefont {Li}, \citenamefont {Kim}, \citenamefont {Fan}, \citenamefont {Wong}, \citenamefont {Tserkovnyak}, \citenamefont {Amiri},\ and\ \citenamefont {Wang}}]{yu2016room}%
  \BibitemOpen
  \bibfield  {author} {\bibinfo {author} {\bibfnamefont {G.}~\bibnamefont {Yu}}, \bibinfo {author} {\bibfnamefont {P.}~\bibnamefont {Upadhyaya}}, \bibinfo {author} {\bibfnamefont {X.}~\bibnamefont {Li}}, \bibinfo {author} {\bibfnamefont {W.}~\bibnamefont {Li}}, \bibinfo {author} {\bibfnamefont {S.~K.}\ \bibnamefont {Kim}}, \bibinfo {author} {\bibfnamefont {Y.}~\bibnamefont {Fan}}, \bibinfo {author} {\bibfnamefont {K.~L.}\ \bibnamefont {Wong}}, \bibinfo {author} {\bibfnamefont {Y.}~\bibnamefont {Tserkovnyak}}, \bibinfo {author} {\bibfnamefont {P.~K.}\ \bibnamefont {Amiri}},\ and\ \bibinfo {author} {\bibfnamefont {K.~L.}\ \bibnamefont {Wang}},\ }\href@noop {} {\bibfield  {journal} {\bibinfo  {journal} {Nano letters}\ }\textbf {\bibinfo {volume} {16}},\ \bibinfo {pages} {1981} (\bibinfo {year} {2016})}\BibitemShut {NoStop}%
\bibitem [{\citenamefont {Labrie-Boulay}\ \emph {et~al.}(2024)\citenamefont {Labrie-Boulay}, \citenamefont {Winkler}, \citenamefont {Franzen}, \citenamefont {Romanova}, \citenamefont {Fangohr},\ and\ \citenamefont {Kl{\"a}ui}}]{labrie2024machine}%
  \BibitemOpen
  \bibfield  {author} {\bibinfo {author} {\bibfnamefont {I.}~\bibnamefont {Labrie-Boulay}}, \bibinfo {author} {\bibfnamefont {T.~B.}\ \bibnamefont {Winkler}}, \bibinfo {author} {\bibfnamefont {D.}~\bibnamefont {Franzen}}, \bibinfo {author} {\bibfnamefont {A.}~\bibnamefont {Romanova}}, \bibinfo {author} {\bibfnamefont {H.}~\bibnamefont {Fangohr}},\ and\ \bibinfo {author} {\bibfnamefont {M.}~\bibnamefont {Kl{\"a}ui}},\ }\href@noop {} {\bibfield  {journal} {\bibinfo  {journal} {Physical Review Applied}\ }\textbf {\bibinfo {volume} {21}},\ \bibinfo {pages} {014014} (\bibinfo {year} {2024})}\BibitemShut {NoStop}%
\bibitem [{\citenamefont {Collins}(2007)}]{collins2007imagej}%
  \BibitemOpen
  \bibfield  {author} {\bibinfo {author} {\bibfnamefont {T.~J.}\ \bibnamefont {Collins}},\ }\href@noop {} {\bibfield  {journal} {\bibinfo  {journal} {Biotechniques}\ }\textbf {\bibinfo {volume} {43}},\ \bibinfo {pages} {S25} (\bibinfo {year} {2007})}\BibitemShut {NoStop}%
\bibitem [{\citenamefont {Horcas}\ \emph {et~al.}(2007)\citenamefont {Horcas}, \citenamefont {Fern{\'a}ndez}, \citenamefont {Gomez-Rodriguez}, \citenamefont {Colchero}, \citenamefont {G{\'o}mez-Herrero},\ and\ \citenamefont {Baro}}]{horcas2007wsxm}%
  \BibitemOpen
  \bibfield  {author} {\bibinfo {author} {\bibfnamefont {I.}~\bibnamefont {Horcas}}, \bibinfo {author} {\bibfnamefont {R.}~\bibnamefont {Fern{\'a}ndez}}, \bibinfo {author} {\bibfnamefont {J.}~\bibnamefont {Gomez-Rodriguez}}, \bibinfo {author} {\bibfnamefont {J.}~\bibnamefont {Colchero}}, \bibinfo {author} {\bibfnamefont {J.}~\bibnamefont {G{\'o}mez-Herrero}},\ and\ \bibinfo {author} {\bibfnamefont {A.}~\bibnamefont {Baro}},\ }\href@noop {} {\bibfield  {journal} {\bibinfo  {journal} {Review of scientific instruments}\ }\textbf {\bibinfo {volume} {78}} (\bibinfo {year} {2007})}\BibitemShut {NoStop}%
\bibitem [{\citenamefont {Lopez-Polin}\ \emph {et~al.}(2022)\citenamefont {Lopez-Polin}, \citenamefont {Aramberri}, \citenamefont {Marques-Marchan}, \citenamefont {Weintrub}, \citenamefont {Bolotin}, \citenamefont {Cerd{\'a}},\ and\ \citenamefont {Asenjo}}]{lopez2022high}%
  \BibitemOpen
  \bibfield  {author} {\bibinfo {author} {\bibfnamefont {G.}~\bibnamefont {Lopez-Polin}}, \bibinfo {author} {\bibfnamefont {H.}~\bibnamefont {Aramberri}}, \bibinfo {author} {\bibfnamefont {J.}~\bibnamefont {Marques-Marchan}}, \bibinfo {author} {\bibfnamefont {B.~I.}\ \bibnamefont {Weintrub}}, \bibinfo {author} {\bibfnamefont {K.~I.}\ \bibnamefont {Bolotin}}, \bibinfo {author} {\bibfnamefont {J.~I.}\ \bibnamefont {Cerd{\'a}}},\ and\ \bibinfo {author} {\bibfnamefont {A.}~\bibnamefont {Asenjo}},\ }\href@noop {} {\bibfield  {journal} {\bibinfo  {journal} {ACS Applied Energy Materials}\ }\textbf {\bibinfo {volume} {5}},\ \bibinfo {pages} {11835} (\bibinfo {year} {2022})}\BibitemShut {NoStop}%
\bibitem [{\citenamefont {Vansteenkiste}\ \emph {et~al.}(2014)\citenamefont {Vansteenkiste}, \citenamefont {Leliaert}, \citenamefont {Dvornik}, \citenamefont {Helsen}, \citenamefont {Garcia-Sanchez},\ and\ \citenamefont {Van~Waeyenberge}}]{vansteenkiste2014design}%
  \BibitemOpen
  \bibfield  {author} {\bibinfo {author} {\bibfnamefont {A.}~\bibnamefont {Vansteenkiste}}, \bibinfo {author} {\bibfnamefont {J.}~\bibnamefont {Leliaert}}, \bibinfo {author} {\bibfnamefont {M.}~\bibnamefont {Dvornik}}, \bibinfo {author} {\bibfnamefont {M.}~\bibnamefont {Helsen}}, \bibinfo {author} {\bibfnamefont {F.}~\bibnamefont {Garcia-Sanchez}},\ and\ \bibinfo {author} {\bibfnamefont {B.}~\bibnamefont {Van~Waeyenberge}},\ }\href@noop {} {\bibfield  {journal} {\bibinfo  {journal} {AIP advances}\ }\textbf {\bibinfo {volume} {4}} (\bibinfo {year} {2014})}\BibitemShut {NoStop}%
\bibitem [{\citenamefont {Winkler}\ \emph {et~al.}(2021{\natexlab{a}})\citenamefont {Winkler}, \citenamefont {Litzius}, \citenamefont {De~Lucia}, \citenamefont {Wei{\ss}enhofer}, \citenamefont {Fangohr},\ and\ \citenamefont {Kl{\"a}ui}}]{winkler2021skyrmion}%
  \BibitemOpen
  \bibfield  {author} {\bibinfo {author} {\bibfnamefont {T.~B.}\ \bibnamefont {Winkler}}, \bibinfo {author} {\bibfnamefont {K.}~\bibnamefont {Litzius}}, \bibinfo {author} {\bibfnamefont {A.}~\bibnamefont {De~Lucia}}, \bibinfo {author} {\bibfnamefont {M.}~\bibnamefont {Wei{\ss}enhofer}}, \bibinfo {author} {\bibfnamefont {H.}~\bibnamefont {Fangohr}},\ and\ \bibinfo {author} {\bibfnamefont {M.}~\bibnamefont {Kl{\"a}ui}},\ }\href@noop {} {\bibfield  {journal} {\bibinfo  {journal} {Physical Review Applied}\ }\textbf {\bibinfo {volume} {16}},\ \bibinfo {pages} {044014} (\bibinfo {year} {2021}{\natexlab{a}})}\BibitemShut {NoStop}%
\bibitem [{\citenamefont {De~Lucia}\ \emph {et~al.}(2016)\citenamefont {De~Lucia}, \citenamefont {Kr{\"u}ger}, \citenamefont {Tretiakov},\ and\ \citenamefont {Kl{\"a}ui}}]{de2016multiscale}%
  \BibitemOpen
  \bibfield  {author} {\bibinfo {author} {\bibfnamefont {A.}~\bibnamefont {De~Lucia}}, \bibinfo {author} {\bibfnamefont {B.}~\bibnamefont {Kr{\"u}ger}}, \bibinfo {author} {\bibfnamefont {O.~A.}\ \bibnamefont {Tretiakov}},\ and\ \bibinfo {author} {\bibfnamefont {M.}~\bibnamefont {Kl{\"a}ui}},\ }\href@noop {} {\bibfield  {journal} {\bibinfo  {journal} {Physical Review B}\ }\textbf {\bibinfo {volume} {94}},\ \bibinfo {pages} {184415} (\bibinfo {year} {2016})}\BibitemShut {NoStop}%
\bibitem [{\citenamefont {Winkler}\ \emph {et~al.}(2021{\natexlab{b}})\citenamefont {Winkler}, \citenamefont {Litzius}, \citenamefont {{De Lucia}}, \citenamefont {Leutner}, \citenamefont {Ramst\"ock}, \citenamefont {Fangohr},\ and\ \citenamefont {Kl\"aui}}]{MicMag2}%
  \BibitemOpen
  \bibfield  {author} {\bibinfo {author} {\bibfnamefont {T.~B.}\ \bibnamefont {Winkler}}, \bibinfo {author} {\bibfnamefont {K.}~\bibnamefont {Litzius}}, \bibinfo {author} {\bibfnamefont {A.}~\bibnamefont {{De Lucia}}}, \bibinfo {author} {\bibfnamefont {K.}~\bibnamefont {Leutner}}, \bibinfo {author} {\bibfnamefont {M.}~\bibnamefont {Ramst\"ock}}, \bibinfo {author} {\bibfnamefont {H.}~\bibnamefont {Fangohr}},\ and\ \bibinfo {author} {\bibfnamefont {M.}~\bibnamefont {Kl\"aui}},\ }\href {https://github.com/micmag2/micmag2} {\bibinfo {title} {{MicMag2}}} (\bibinfo {year} {2021}{\natexlab{b}})\BibitemShut {NoStop}%
\end{thebibliography}%

\section*{Acknowledgements}

EMJ acknowledges the “Alexander von Humboldt Postdoctoral Fellowship”. Support from DFG (Priority Program “Skyrmionics”), the the European Research Council (ERC) under the European Union's Horizon 2020 research and innovation program (Grant No. 856538, project “3D MAGiC”)" and Horizon Europe project NIMFEIA (101070290) are acknowledged. \newline

\section*{Author contributions statement}

E.M.J. conceived the experiments, K.L. carried out the theoretical and micromagnetic calculations, E.M.J., M.G.F. and J.M.M. conducted the experiments, E.M.J., K.L. and R.F. discussed the results. All authors reviewed and contributed to the writing of the manuscript. The project was supervised by R. F. and M. K.

\section*{Competing interests}
The authors declare no competing interests.
The corresponding author is responsible for submitting a \href{http://www.nature.com/srep/policies/index.html#competing}{competing interests statement} on behalf of all authors of the paper. This statement must be included in the submitted article file.

\end{document}